\documentclass[12pt]{iopart}

\usepackage{iopams}  
\usepackage{hyperref}
\usepackage{graphicx}
\usepackage{bm}
\usepackage{mathptmx}
\usepackage[utf8]{inputenc}
\usepackage{xcolor}
\usepackage{calrsfs}
\DeclareMathAlphabet{\pazocal}{OMS}{zplm}{m}{n}
\begin{document}

\title[]{Application of interpretable machine learning for cross-diagnostic inference on the ST40 spherical tokamak}
\author{T. Pyragius, C. Colgan, H. Lowe, F. Janky, M. Fontana, Y. Cai, G. Naylor, and the ST40 team}
\address{Tokamak Energy Ltd., 173 Brook Dr, Milton, Abingdon, OX14 4SD, UK}
\ead{tadas.pyragius@tokamakenergy.co.uk}
\vspace{10pt}
\begin{indented}
\item[]July 2024
\end{indented}

\begin{abstract}
Machine learning models are exceptionally effective in capturing complex non-linear relationships of high-dimensional datasets and making  accurate predictions. However, their intrinsic ``black-box'' nature makes it difficult to interpret them or guarantee ``safe behavior'' when deployed in high-risk applications such as feedback control, healthcare and finance. This drawback acts as a significant barrier to their wider application across many scientific and industrial domains where the interpretability of the model predictions is as important as accuracy. Leveraging the latest developments in interpretable machine learning, we develop a method to parameterise ``black-box'' models, effectively transforming them into ``grey-box'' models. We apply this approach to plasma diagnostics by creating a parameterised synthetic Soft X-Ray imaging $-$ Thomson Scattering diagnostic, which predicts high temporal resolution electron temperature and density profiles from the measured soft X-ray emission. The ``grey-box'' model predictions are benchmarked against the trained ``black-box'' models as well as a diverse range of plasma conditions. Our model-agnostic approach can be applied to various machine learning architectures, enabling direct comparisons of model interpretations.
\end{abstract}

%
%
%
%
%

\section{Introduction}
Advancements in plasma physics, particularly in the context of fusion energy research, have significantly benefited from the application of machine learning techniques (see reviews \cite{plasma_data_science_review, plasma_ml_and_bayesian_review}). These methods have proven effective in deciphering complex data generated from fusion experiments, where traditional analytic approaches can fall short. A key challenge in this domain is the interpretation of results from machine learning models, especially when these models function as ``black boxes'', providing little insight into the underlying physical phenomena they model. A notable exception to this are physics informed neural networks PINNs, where the equations governing the dynamics of the system constrained by laws of physics are embedded in the loss function of the machine learning model \cite{pinns}. With PINNs the model is fully interpretable because it necessarily satisfies the physics constraints. However, PINN use cases impose additional limits on their wider adoption as a new dataset input requires complete model retraining. Moreover, PINN approach is not always possible, especially if the underlying laws are unknown or are prohibitively expensive to model. In these situations a different approach is required where attempts are made to extract the underlying mechanisms learned by the model from the data. \\ \\
A lack of model interpretability has several disadvantages. First, unlike traditional physics models that provide an analytical framework to explain interactions between physical observables and their impacts on measured outcomes, ML models offer predictions without explaining the mechanics behind them. Consequently, it becomes very difficult to discern the reasons for these predictions, thereby impeding our understanding of the governing physical mechanisms. This lack of transparency directly affects our ability to understand the underlying physics and most importantly design experiments to falsify the model and find its limitations. This issue not only erodes trust in machine learning models but also restricts their wider adoption in physics, where knowing the relationships between experimental observables is essential. \\ \\ 
In the context of plasma physics and control, the interpretability and regions of model validity are of significant importance for mission critical applications and their reliable deployment for real-time inference and plasma control applications \cite{attempt_at_explainable_ai_for_plasma, keras2c_paper, keras2c, machine_learning_pcs, rt_profile_prediction, fpga_ml}. Conventionally, such control is achieved via a  combination of diagnostic measurements e.g. magnetic field measurements, density etc., and feedback control actuation derived from them. However, there are inherent trade-offs in diagnostics that perform these measurements. These trade-offs range from limited time and spatial resolution, dynamical range as well as the cut-off limits imposed by the plasma conditions under which the diagnostic operates. In addition, whilst some diagnostics may provide measurements that benefit from high temporal resolution, the results may not be trivial to interpret. Consequently, developments in creating ``synthetic diagnostics'' using machine learning methods have enabled offsetting the inherent trade-offs present in these diagnostics where their desirable properties can be combined \cite{synthetic_diagnostic_example_1,synthetic_diagnostic_example_2, synthetic_diagnostic_example_3,synthetic_diagnostic_example_4}. However, this comes at the expense of understanding the underlying mechanics that drive the predictions, making their adoption for machine and plasma control unappealing.  \\ \\
Inspired by previous work \cite{synthetic_diagnostic_example_3, sxr_temperature_multi_energy, sxr_temperature_reconstruction_analytic}, we propose a methodology that leverages recent advances in state-of-the-art interpretable machine learning methods that enable the transition away from ``black-box'' machine learning models to ``grey-box models". Using a combination of Accumulated Local Effects (ALE) \cite{ale_main, ale_py, ale_code, alibi, alibi_git}, Shapley Additive exPlanations (SHAP) \cite{shap_original, shap_comp, shap_code} and Symbolic Regression \cite{sr, py_sr} techniques we show how local and global explanations for model predictions can be successfully parametrised to a closed functional form without experiencing significant degradation in their predictive performance. As a case study, we develop a synthetic ``grey-box'' diagnostic that provides high temporal and spatial resolution electron temperature and density predictions from the measured soft X-ray (SXR) signals. We show how the parametrised model gives insight into individual contributions of input emission towards the predicted electron temperature and density on a local and global scale of measured x-ray signals. This directly reveals feature importance of each line of sight and their consequent contribution to the predicted output. Understanding the input feature importance and their evolution enables us to reduce the ``black-box'' model complexity to fewer model parameters making the model behaviour fully predictable and understandable to humans. This further provides added corroboration as we are able to determine how the model behaves across different scenarios allowing us to determine regions of interest where the model produces reliable predictions and where it does not.\\ \\
Finally, the methodology and techniques applied in this work are model-agnostic and simple to implement on other fusion-relevant diagnostics. This makes our approach highly accessible and easily generalisable to various supervised machine learning architectures, including Feed-Forward Neural Networks (FFNN), Gaussian Process Regression (GPR), and ensemble Random Forests (RF). Given that each model architecture runs on different principles and optimisation routines, we show how this leads to different results in model interpretation. \\ \\ 
This paper is organised as follows. In Section 2 we give a brief description of our diagnostic setup. In section 3 we outline our data preparation workflow that we use to pre-process data for model training and validation. Additionally, we discuss the tools used for model interpretation and how they can be used to parametrise a machine learning model. In section 4 we present the results of the different model performance and their interpretation. In the final section we draw conclusions and discuss some potential future work derived from the results presented in this work.

\section{Description of the diagnostic setup}
\begin{figure*}[t!]
\includegraphics[width=\textwidth]{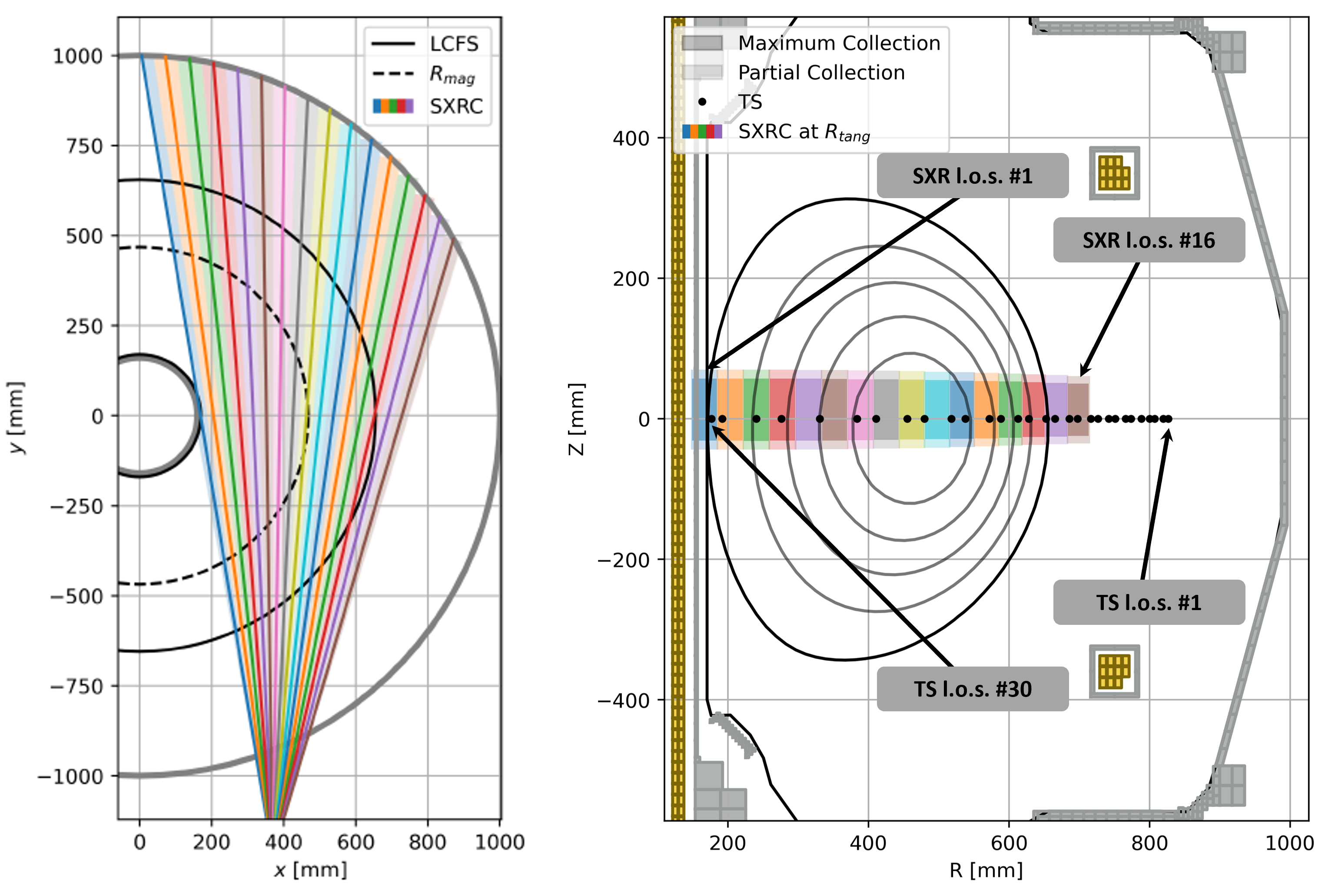}
\put(-430,290){(a)}
\put(-250,290){(b)}
\caption{\label{fig:diagnostic_layout} Diagnostic layout (a) top view of the tokamak (b) cross-section of the tokamak. Here the high-field side corresponds to $R<0.35$~m, plasma core $0.35<R<0.55$~m and the low-field side to $R>0.55$~m. The low field side soft X-ray line of sight (los) 16 does not cross the core of the plasma – low temperature /density. For the Thomson Scattering diagnostic the low field side line of sight number is 1. The SXRC is positioned inside the inner vacuum chamber of ST40, with a tangential view across the plasma. We observe that the first 10 lines of sight on the low field side of the TS do not overlap with SXRC \cite{sxr_cary}.}
\end{figure*}

The diagnostics and the measured experimental data discussed in this work were obtained on the high-field spherical tokamak, ST40, operated by Tokamak Energy Ltd (see \cite{st40} for machine details). Here, we will give a brief description of each diagnostic. The measured soft X-ray (SXR) signal emission from the plasma is provided by the Soft X-Ray Camera (SXRC). It consists of 16 channels viewing tangentially across the plasma (see figure~\ref{fig:diagnostic_layout}). The emission signals are captured with an acquisition rate of 500~kHz (2~$\mu$s) which is further low-pass filtered to 4~kHz (0.25~ms) to remove the high frequency noise from the signal. 
The diodes are protected by a 25~$\mu$m-thick Be filter, so that the SXRC system can reliably detect energies between 1~keV and 10~keV (for further details, see \cite{sxr_cary}).
The local emissivity of the plasma is given by 

\begin{figure*}[t!]
\vspace{-1.cm}
\centering
\includegraphics[width=0.7\textwidth]{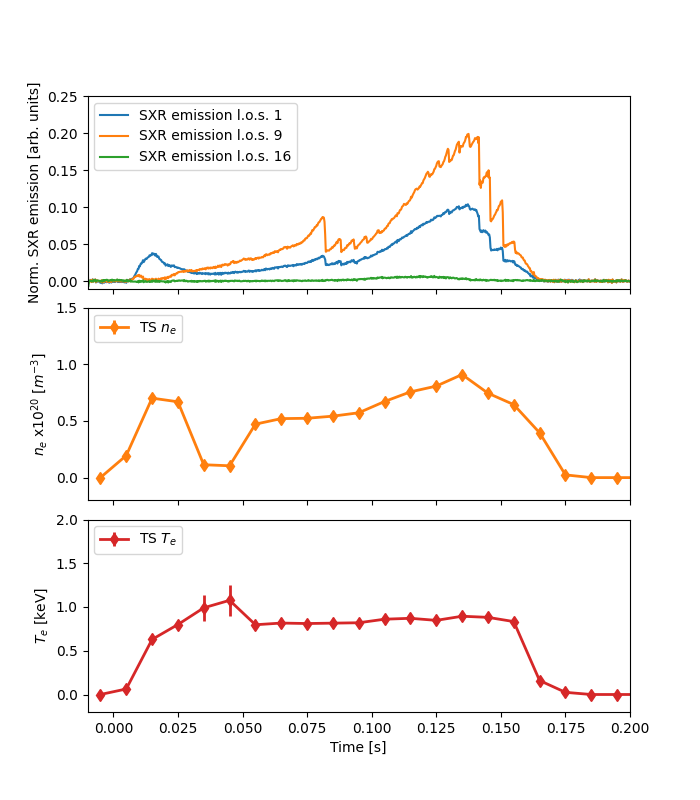}
\put(-320,310){(a)}
\put(-320,215){(b)}
\put(-320,120){(c)}
\vspace{-1cm}
\caption{\label{fig:sxr_ts_data} ST40 plasma pulse 11574. (a) A set of measured soft X-ray emissivities for three different lines of sight. The first line of sight $\#$1 is the closest to the high field side in the RZ plane field of view (see figure~\ref{fig:diagnostic_layout} for details). The line of sight $\#$9 roughly crosses the core of the plamsa in the same RZ plane. The line of sight $\#$16 is the most low field side line of sight looking on the edge of the plasma which does not cross the core. (b) and (c) time evolution of TS density and temperature respectively of line of sight 24 ($R=0.45$~m).}
\end{figure*}

%
\begin{equation}
       \epsilon_{\textrm{local}} = n_e \sum_{Z} L_{Z} n_{Z} \ ,
\end{equation}
where $L_{Z}$ is the cooling factor, $n_z$ is the density of ion species $Z$, and $n_e$ is the electron density. For X-ray emission energies far greater than the characteristic radiation of the ion, the cooling factor is purely from bremsstrahlung radiation. This is the case for the radiation seen by the SXRC diagnostic from the hydrogen fuel, as well as from carbon impurities from the walls of ST40. However, across the temperatures and densities measured here, the cooling factors of heavier ions are orders of magnitude greater due to line radiation, with non-monotonic dependencies on $T_e$ and impurity transport \cite{putterich2019determination}.
Contributions from line radiation to the signals on the SXRC diagnostic have already been observed on ST40 pulses, due to the presence of trace heavy impurities from diagnostic gases, the molybdenum divertor plates and the in-vessel magnetic coils used for startup \cite{sxr_cary}.
Accurately calculating the atomic spectra required for these cooling factors across the many impurity species, charge states and the range of plasma parameters is a challenging issue for current fusion devices.
Similarly, accurately diagnosing the full plasma composition and its evolution requires an extensive suite of radiation measurements and analysis.
Hence, previous applications of machine-learning labelled by Thomson Scattering (TS) measurements have focused on removing contributions of line radiation to multi-energy soft X-ray signals, to provide electron temperature predictions \cite{synthetic_diagnostic_example_3, sxr_temperature_multi_energy}.
Instead here, we focus on a single-filtered, spatially resolving soft X-ray diagnostic to provide estimations of the profiles of both $T_e$ and $n_e$ to high temporal resolution.
From the local emissivity, the emissions, $\epsilon$,  measured by the SXRC are given by integrating along the lines of sights of the diagnostic,  
\begin{equation}
    \epsilon = \int \epsilon_{\textrm{local}} \ dl \ .\label{eq:emission}
\end{equation}
Figure~\ref{fig:sxr_ts_data} (a) shows a typical SXR emission evolution for different lines of sight during a plasma discharge. The measurement of localised electron temperature and density is provided by the Thomson Scattering (TS) diagnostic. The TS system consists of Nd:YAG laser with 1~J output energy running at 100~Hz (10~ms) repetition rate. The scattered Doppler shifted light from the thermal electrons in the plasma is gathered by the collection optics with 30 spatial locations across the plasma mid-plane core and pedestal with the radial size of the scattering region varying from 7~mm in the outboard to 36~mm on the inboard (for further details see ref.~\cite{ts_hazel}). Because of the positioning of the TS in relation to the SXRC diagnostic, the first 10 low-field side lines of sight of the TS do not overlap, see figure~\ref{fig:diagnostic_layout}. Figure~\ref{fig:sxr_ts_data} (b) and (c) show a typical TS electron density and temperature evolution for given line of sight during a plasma discharge which we compare to the SXR emission. \\

\hspace*{-0.85cm}The main motivation behind developing a synthetic diagnostic is that it enables an efficient extraction of favourable qualities from a chosen set of diagnostics whilst neglecting the unfavourable ones. In our case the SXRC diagnostic provides a high temporal resolution measurement of SXR emission. However, these measurements are difficult to interpret due to their complex physical origin (lack of knowledge of impurities and their transport) as well as the fact that the measurements are line integrated. On the other hand, the TS diagnostic yields localised measurements of electron temperature and density which are easy to interpret. However, the drawback is that the temporal resolution of the measurement is constrained by the laser repetition rate. Using a machine learning framework, we are able to distill the favourable qualities of each diagnostic whereby the high temporal resolution SXR measurements can be used in combination with the TS profile measurements providing a high temporal resolution and localised electron temperature and density predictions. This is achieved by the machine learning model ``learning'' a mapping between the measured emission data and the corresponding TS electron temperature and density. However, this ``learned'' mapping is not typically accessible from the models and therefore cannot be interpreted for human understanding.

\section{Description of the machine learning setup}
\subsection{Data preparation workflow}
We briefly outline our data preparation workflow. Our training dataset consists of 198 valid plasma pulses containing both soft X-ray (SXR) and Thomson Scattering (TS) signals. The plasma scenarios within the set cover a wide range of plasma conditions including hot-ion, H-mode, ohmic, neutral-beam injection (NBI) heated, diverted, low impurity, MHD, limited high-beta to name just a few. We do not specify the precise conditions of the plasma scenario to the model; instead, we input signals of SXR emission, TS temperature, and density. First, the SXR signals are filtered using low-pass filtering techniques to remove high frequency electronic noise. We perform additional filtering on the measured TS signals as follows; during the campaign the measured electron temperatures did not exceed 2.5~keV (constrained by the plasma scenarios). However, under some conditions fitting of the TS signals to extract temperature and density estimates is unreliable. This occasionally leads to unphysical temperature and density estimates (e.g. 10 keV temperature). These spurious measurements normally occur at the start and the end of the plasma pulse (e.g. due to presence of mixed impurities) where the SXR emission signals are close to zero. To address this, we perform simple threshold based filtering to remove these unphysical values and set them to zero, such that for zero SXR emission signals we have zero temperature/density. We optionally pad the train/test data sets with zeros in order to ensure that for zero TS density and temperature we have zero SXR signal. We find that this approach helps remove the DC offset present in the predicted temperature and density values after training. \\ \\
The measured TS and SXRC diagnostic signals cover the same time period, however, their temporal resolutions differ significantly between them with SXRC having a very high temporal resolution measurement over the TS. As a result, to align these two sets of measurements and compare them effectively, we identify points in time where data is available for both measurements. Specifically, we locate instances where the timestamps of the TS measurements match with those of the downsampled (binned) and filtered SXRC measurements. These matching points in time are used in both training and testing datasets. We remove the remaining unmatched data points from SXR data which constitute 98\% of the data discarded. \\ \\ 
The processed dataset consists of a time-ordered sequence of matched SXR-TS pairs for each plasma pulse. We sequentially concatenate these batches of data for each plasma pulse. The resultant dataset is a time-ordered sequence of these plasma pulses. Next, we intentionally scramble the entire dataset while preserving the temporal structure within each SXR-TS pair. This process disrupts the time-ordered sequence within each plasma pulse as well as between different pulses. \\ \\ 
The last stage of data processing involves normalisation of the SXR and TS signals. Here the input SXR signals are normalised to be in the range $[0, 1]$ which is a requirement for Feed-Forward Neural Networks (FFNN). This requirement, however, does not need to be satisfied for Random Forest (RF) or Gaussian Process Regression (GPR) based machine learning models which can deal with un-normalised data. Nevertheless, for consistency, we use the same normalised inputs for these models. The output features are measured TS electron temperatures and densities in units of [keV] and $[m^{-3}/10^{20}]$ respectively. \\ \\
The final cleaned dataset contains approximately 7k of input-output pairs of data points. The model performance is validated against a test dataset consisting of 20\% (1.4k input-output pairs) randomly drawn from the train / test split. We exclude certain plasma pulses from the training/testing data for additional validation. 


\subsection{Model architectures}
In this work we apply interpretability tools on three common model architectures used in supervised machine learning: Feed-Forward Neural Networks (FFNN) \cite{tensorflow}, Random Forests \cite{rf_gpr} (RF), and Gaussian Process Regression (GPR) \cite{rf_gpr}. All models are commonly used for regression based tasks and are very capable of capturing complex non-linear relationships and their interactions present in multi-dimensional datasets. Moreover, they are all compatible with the interpretability tools enabling direct comparison and bench-marking between them. The underlying motivation to use multiple models for the same problem i.e. building a synthetic diagnostic is multifaceted. Firstly, each model brings unique strengths and weaknesses to the table, allowing for a more comprehensive understanding of the problem at hand. This aspect is particularly enhanced when combined with the interpretability tools as it gives direct access to the reasoning behind each model's decision and whether it is compatible with physics driven expectations. For example, GPR is ideal when training data is both costly and limited in quantity. Furthermore, it has inbuilt uncertainty quantification which is excellent for guiding our trust in the predicted outputs. However, GPR models are not inherently suitable for deployment on hardware for real-time applications, which restricts their use as real-time synthetic diagnostics. FFNNs are on the other hand readily deployable on hardware via keras2c, hls4ml frameworks or other compilers e.g. MATLAB Simulink making them ideal for real-time applications in fusion science \cite{keras2c_paper, keras2c, machine_learning_pcs, rt_profile_prediction, fpga_ml,  fastml_paper, fastml_code, real_time_control}. However, unlike GPR, FFNNs do not provide easy to obtain and reliable uncertainty quantification measure of the model predictions and typically require large training datasets to obtain good performance. In certain areas of comparison RFs strike the middle ground between FFNNs and GPRs. When compared to FFNNs, RFs are generally more robust to overfitting and can be extended to include error estimation via quantile regression. This list is non-exhaustive, but illustrates important differences between the models, their potential scope of applications and interpretation. \\ \\
All models contain 16 inputs of normalised SXR emission measurements and 60 outputs containing normalised TS temperature and density measurements for all 30 lines of sight. We additionally train models with a single output feature for a given line of sight to predict temperature or density. The rationale behind this is to provide a simple and easy to interpret set of examples and demonstrate how the explainability tools scale. \\ \\ 
We build and train our models using well established open source libraries scikit-learn \cite{rf_gpr} for Random Forests and Gaussian Process regression and TensorFlow \cite{tensorflow} for Feed-Forward Neural Networks. For our GPR model, we use a radial basis function (RBF) kernel (also known as squared exponential). Our Random Forest consists of $n=100$ estimators and is wrapped in mulit-output regressor to enable multi-output predictions. The rest of the tunable hyperparameters in GPR and RF models are default as preset by the libraries used. We do not perform any additional hyperparameter tuning. The FFNN is a two layer deep fully connected network. The first deep layer contains 64 nodes, while the second layer contains 32 nodes. We use a Rectified Linear Unit (ReLU) activation function for both layers. The models is trained on ``Adam'' optimiser \cite{adam} with a mean-squared error loss function. All of the trained models evaluated on the test dataset show similar mean-absolute error (MAE) performance (see Table~\ref{tab:mae_performance} for model performance summary). \\ \\
\begin{figure*}[t!]
\vspace{-0.5cm}
\centering
\includegraphics[width=0.8\textwidth]{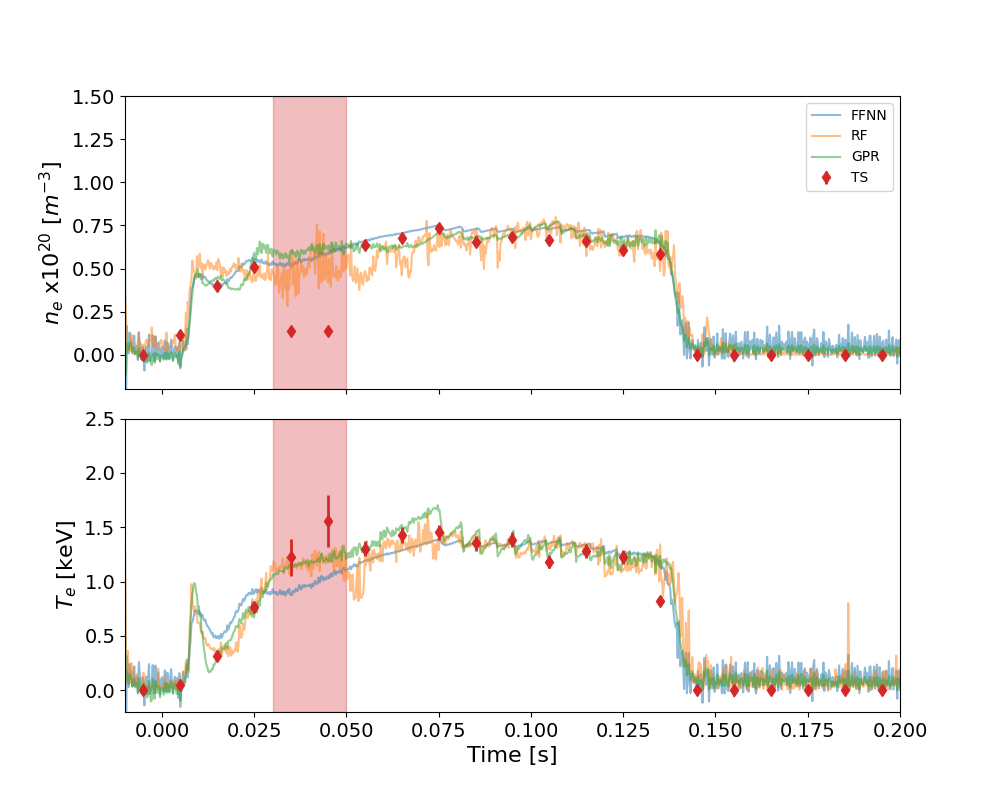}
\put(-350,250){(a)}
\put(-350,135){(b)}
\vspace{-0.5cm}
\caption{\label{fig:model_predictions} ST40 plasma pulse 11573. Time evolution of the electron density (a) and temperature (b) as predicted by the different ML models benchmarked against measurements made by the Thomson Scattering diagnostic for radial position $R=0.53$~m (line of sight 20). }
\end{figure*}
\begin{figure*}[t!]
\vspace{-0.5cm}
\centering
\includegraphics[width=0.75\textwidth]{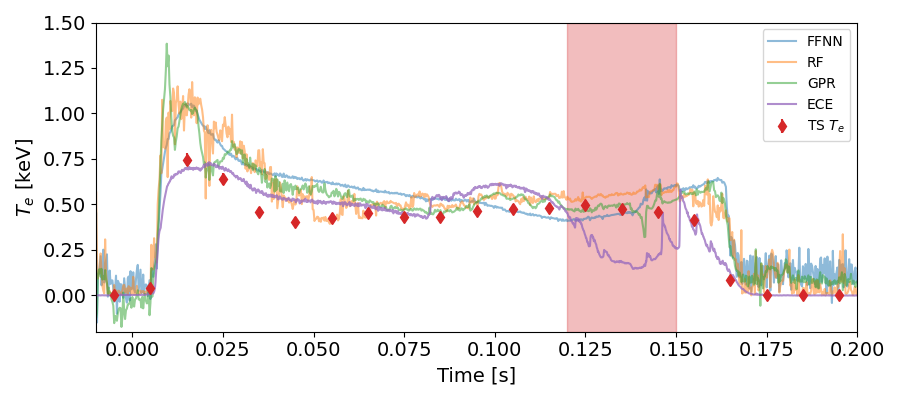}
\vspace{-0.5cm}
\caption{\label{fig:ece_comparison} ST40 plasma pulse 11574. Time evolution of the electron density temperature as predicted by the different ML models benchmarked against measurements made by the Thomson Scattering and the electron-cyclotron emission (ECE) diagnostics. The TS radial position is $R=0.33$~m and the closest ECE position is $R=0.36\pm0.01$~m. The shaded region in the figure corresponds to where the ECE experiences a cut-off limit yielding unreliable temperature measurements.}
\end{figure*}
\begin{figure*}[t!]
\vspace{-0.5cm}
\centering
\includegraphics[width=0.8\textwidth]{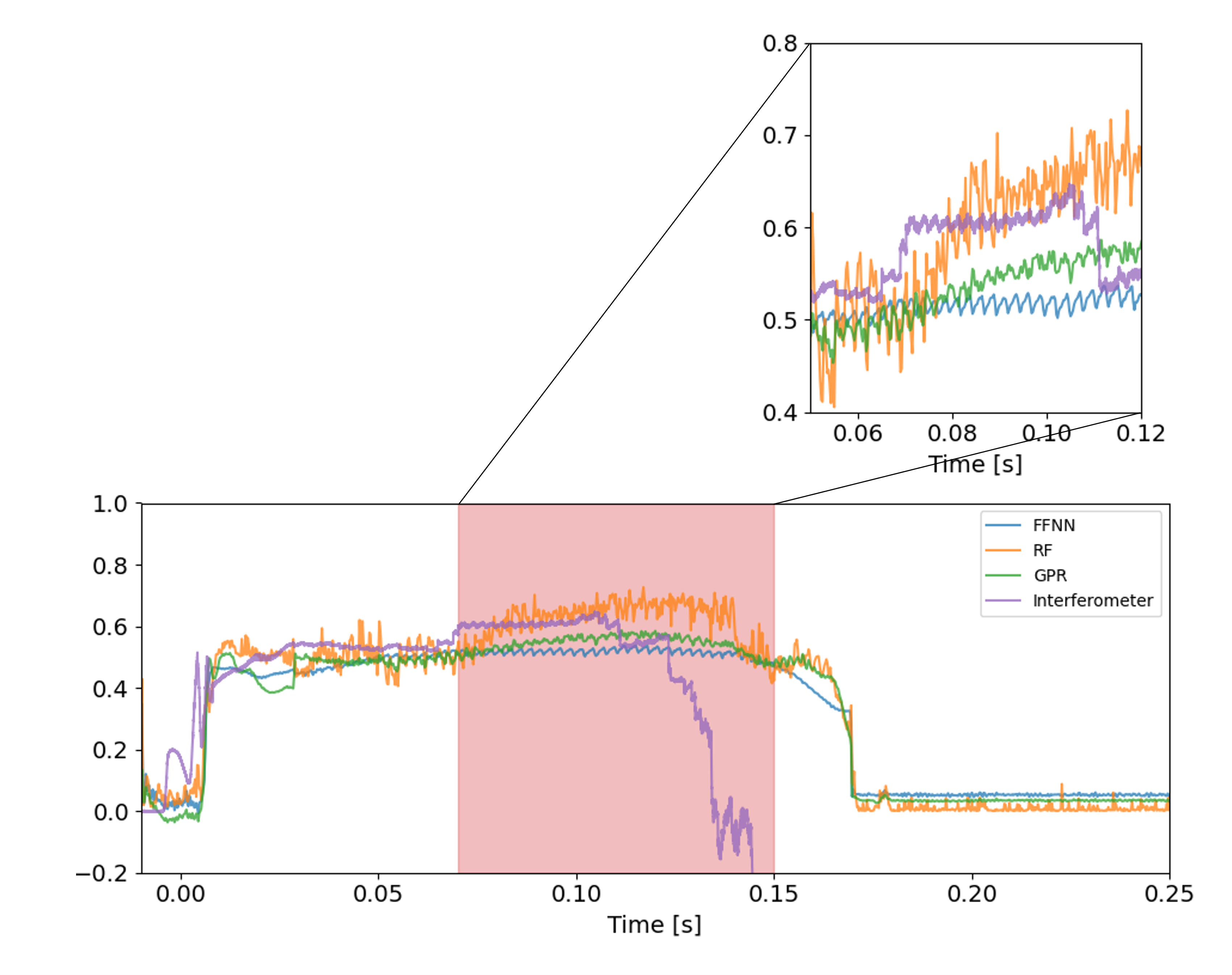}
\put(-155,270){(b)}
\put(-350,135){(a)}
\put(-350,35){\rotatebox{90}{$\langle n_e l\rangle \times 10^{20}$ [$m^{-2}$]}}
\put(-155,175){\small\rotatebox{90}{$\langle n_e l\rangle \times 10^{20}$ [$m^{-2}$]}}
\vspace{-0.5cm}
\caption{\label{fig:interferometer_comparison} ST40 plasma pulse 11805. Time evolution of the integrated electron density (a) as predicted by the different multi-output ML models benchmarked against measurements made by the line integrated density of the sub-millimetre interferometer. The model density is calculated using eq.~\ref{eq:int_density}. Here, we observe sawtooth crashes in line integrated density which are also seen in the model predictions, though to varying degree. (b) is the magnified comparison between the models and the interferometer density during the flattop phase.}
\end{figure*}
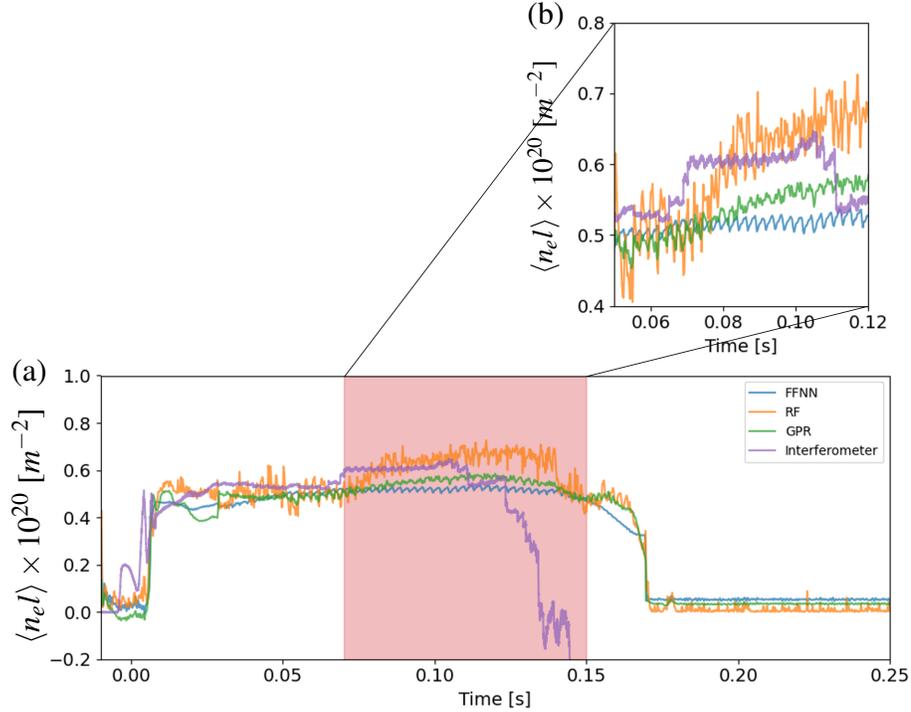
\begin{table}[t!]
\caption{Model performance metrics based on mean-aboslute error estimation on the test dataset. No. of outputs refer to the number of output line of sight pairs the model is capable of predicting for $T_e$ and $n_e$. Each model contains 16 SXRC emission inputs.}\vspace*{0.25cm} 
\centering 
\begin{tabular}{l c c c c c} 
\hline\hline
Model Name & No. of Outputs & MAE $T_e$ [keV] & MAE $n_e$ $[m^{-3}/10^{20}]$\\
[0.5ex]
\hline\hline
FFNN & 30 & 0.15 & 0.07   \\[1ex]
RF & 30 & 0.10 & 0.03  \\[1ex]
GPR & 30 & 0.15 & 0.06 \\[1ex]

\hline 
\end{tabular}\label{tab:mae_performance}
\end{table}

\hspace*{-.8cm}Figure~\ref{fig:model_predictions} shows the predicted electron temperature and density time evolution for a radial position $R=0.53$~m (line of sight 20) based on the SXR measured emissivities. We additionally overlay the measured TS temperature and density. The plasma pulse 11573 was excluded from the training data and has not been seen by any of the models under test. We find the predictions agree reasonably well with the measured TS temperatures and densities. A large deviation between the measurement and the predictions is observed around 25ms and 50ms. This is where the TS diagnostic experiences mechanical vibrations which affect the collection efficiency of scattered light resulting in a systematic dip in the density and an increase in error estimation of the temperature (due to lower SNR). However, the measured SXR input emission is robust to mechanical vibrations, so the model does not reflect these vibrations in the predicted electron temperature and density estimates. We additionally benchmark the predicted electron temperature against our electron cyclotron emission diagnostic (ECE), see figure~\ref{fig:ece_comparison}. The ECE diagnostic provides high temporal resolution and spatially localised measurements of the electron temperature, but has the limitation of working only in a restricted range of density/temperature. These conditions are typically satisfied in the plasma flat top phase which is where the comparison is made. We observe a good agreement between TS, ECE and the predicted temperatures during the flattop phase. All models appear to overestimate the predicted temperatures at the startup and the end of the plasma pulse. The ECE experiences a cutoff limit between 0.13-0.14~s window resulting in spurious temperature measurements which the measured TS and the model predicted temperatures are not susceptible to. Unfortunately, we did not have an interferometer integrated line density measurement for this specific shot. However, we were able to obtain a partial microwave interferometer trace in our most recent campaign, as shown in the figure~\ref{fig:interferometer_comparison}. The interferometer measures line integrated density along a radial midplane \cite{st40}. The predicted multi-output model densities can be summed over to produce line integrated density (analogous to how line integrated density is estimated using TS) and compared directly to the measurements of the interferometer 
\begin{equation}
    \langle n_e l\rangle = 2\sum_i n_i \Delta l_i \label{eq:int_density}
\end{equation}
where $n_i$ is the predicted density for a TS line of sight $i$ along a line element $\Delta l_i$. Here, the factor of 2 in the calculation arises because the interferometer measures the integrated line density by traversing the plasma twice; this occurs as the interferometer beam is reflected back by a mirror on the central column. From the figure the measured signal experiences a phase jump around 0.14~s due to a signal loss during the pulse. During the flattop phase, the measured integrated line density shows small sawtooth crashes which are also observed in the machine learning models to varying magnitude and in phase.  \\ \\
Once the model has been trained and validated against ``unseen'' test data to ensure nominal performance without overfitting, it is typically deployed into production. This means it is integrated into a workflow to make regular predictions based on new input data. In some cases, adversarial examples are constructed to further stress-test the model and identify scenarios where it may fail. However, these models are often ``black-boxes'', meaning their predictions cannot be directly interpreted from the inputs.

\section{Model interpretation}
\subsection{Mutual Information}
Before constructing our machine learning model, understanding the relationship between input and output variables is crucial. While physics equations typically provide closed analytical forms for this purpose, they may not be feasible in complex systems with numerous inputs and unknown interactions \cite{mutual_info}. This challenge is particularly apparent in the study of plasmas, where a large suite of complex diagnostics and analysis techniques are needed to understand plasma temperature, density and their complex dependence on various plasma parameters. One approach that can provide a good initial head start is to calculate their Mutual Information (MI). MI is an information theoretic method that measures the amount of information that can be obtained about one variable by observing another. It is a measure of the mutual dependence between the two variables. Essentially, it quantifies how much knowing one of these variables reduces variance about the other. It is defined as
\begin{eqnarray}
    I(X,Y) = \sum_{y \in Y} \sum_{x \in X}
    { P_{(X,Y)}(x, y) \log\left(\frac{P_{(X,Y)}(x, y)}{P_X(x)\,P_Y(y)}\right) },\label{eq:mi}
\end{eqnarray}
where $P(x,y)$ is the joint probability distribution and $P(x)$ and $P(y)$ are the marginal probability distributions of observables X and Y, respectively. These probability distributions can be obtained by binning the experimental data into histograms and normalising them. When determining dependence between input and output variables mutual information has an advantage over a simple correlation estimates in that it is able to capture both linear and non-linear dependence \cite{mutual_information}. This characteristic feature of MI estimates is ideal in capturing complex non-linear relationships present in high-dimensional data of plasma systems. 

\subsection{Accumulated Local Effects}
Accumulated Local Effects (ALE) are a model agnostic tool used in the interpretability of machine learning models, focusing on understanding the effects of features on the predictions of the model. ALE looks at how changes in a feature affect the predictions on average, locally. Here, the term ``locally'' refers to calculating the effects of a particular feature at specific points or intervals in the feature space, rather than considering the global or overall effect of a feature across its entire range. This method is particularly useful for identifying and interpreting complex interactions between features and the output predictions across different range of the feature space (see \cite{ale_main, ale_py, ale_code, alibi, alibi_git} for further examples). For a prediction of a model $f(x)$ (where $x$ is a vector of input features) the ALE of a feature $x_i$ quantifies the impact of $x_i$ on the prediction $f(x)$, isolating its effect from other features by averaging out their influences. This is calculated by first partitioning the range of feature values $x_i$ to $k$ intervals. For each interval of length $l$ of feature $x_i$ the difference in the model's prediction when $x_i$ changes within this interval is calculated whilst keeping other features constant. This difference is then averaged over all instances that fall into the interval. This is given by
\begin{equation}
    ALE_l = \frac{1}{n_l} \sum_{x \in l} \left[ f(x_{-i}, x_{i}^{high}) - f(x_{-i}, x_{i}^{low}) \right], \label{eq:ale_i}
\end{equation}
where $n_l$ is the number of instances in interval $l$, $x_{-i}$ represents the feature vector excluding $x_i$, and $x_{i}^{high}$ and $x_{i}^{low}$ are the upper and lower bounds of the interval $l$ for feature $x_i$. To determine the interval $l$ we use the Freedman-Diaconis rule which uses the interquartile range (IQR) to determine the bin width, rather than the standard deviation. This makes it more robust to outliers since the IQR is less sensitive to extreme values than the standard deviation \cite{freedman_diaconis}. \\ \\ 
The estimated ALEs can be used to provide a functional decomposition, meaning that adding all ALE estimates yields the prediction function \cite{ale_main, explainable_ai_book_2}. Under conditions where the estimated ALEs are dominated by individual features and their first order interactions (i.e. any given pair of input feature interactions), the ``black-box'' model can be approximated to
\begin{equation}
    f(x)\approx f_0+ \sum_{i=1}^{n}ALE(x_i)+ \sum_{1\leq i< j \leq n}^n ALE(x_i,x_j), \label{eq:ale}
\end{equation}
where $f_0$ is constant offset and $ALE(x_i, x_j)$ describe the first order interaction effects. For $n$ input features we have $C(n,2)$ combinations of pair-wise interaction effects to compute. As a result, the expression given in eq.~(\ref{eq:ale}) can be decomposed as a sum of its functional components  \cite{ale_main, explainable_ai_book_2}
\begin{equation}
    f(x)\approx f_1(x_1)+ f_2(x_2)+...+f_n(x_n) + \sum_{1\leq i< j \leq n}^n f_{i,j}(x_i,x_j). \label{eq:parametric}
\end{equation}
This is known as the Hoeffding-Sobol decomposition or the ANOVA representation (Analysis of Variance) \cite{anova_0, anova_1, hoeffding_sobol}. It allows us to approximate what the machine learning model has learned by transforming it into a parametric model that captures low-order interaction effects. In this decomposition, we have neglected higher-order interaction terms by further assuming that the main contributions to the model predictions come from low-order interactions \cite{rabitz}. 
\\ \\
ALE estimates are computationally fast and can be rapidly generated for quick model interpretation. Since ALE calculates the average effect of a feature over a chosen interval range, it can potentially miss more intricate behavior captured by the model. For instance, the computed ALE for a given feature may show a flat monotonic trend over the input feature value range. Yet, if the feature's contribution to the prediction has a bifurcated relationship depending on another feature value (e.g., a binary value or some threshold), this relationship would be averaged out and unnoticed. Similarly, the feature contribution may exhibit oscillatory behavior, which may be averaged out if the interval size $l$ is larger than the periodicity of the feature effect. In principle, this could be circumvented by choosing a smaller interval size. However, sparse, inhomogeneously distributed, or particularly noisy data can produce spurious estimates. Consequently, ALE may not always provide a reliable interpretation of a chosen model and dataset. To further enhance model interpretability, we turn to a different approach that addresses some of the issues associated with ALE.

\subsection{Shapley additive explanations (SHAP)}
Shapley additive explanations is a game theoretic approach that attempts to explain the output of a machine learning model. It provides a method to fairly distribute the ``payout'' (the prediction of a model $f(x)$) among the ``players'' (the features $x_i$ of the model). For each feature $x_i$ contained in the feature set $N$, the Shapley value $\phi_i$ quantifies the contribution of feature $i$ to the difference between the actual prediction $f(x)$ and the average prediction over the dataset, namely \cite{shap_original, shap_nature}:
\begin{equation}
    \phi_i = \sum_{S \subseteq N \setminus \{i\}} \frac{|S|!(N-|S|-1)!}{N!} \left[f(S \cup \{i\}) - f(S)\right],\label{eq:shap_i}
\end{equation}
where $S$ is a subset of features excluding $i$, $|S|$ is the number of features in subset S, $f(S)$ is the model prediction without feature $i$, $f(S \cup \{i\})$ is the prediction with feature $i$, and $N$ is the total number of features. The factorial term calculates the weight of each subset $S$, accounting for all permutations of features being added in any order and the $\left[f(S \cup \{i\}) - f(S)\right]$ represents the marginal contribution of feature $i$ when added to subset $S$.  SHAP can be extended to determine interaction effects among any pairs of ``players'' (the features $x_i$ and $x_j$ of the model) \cite{shap_code, shap_nature, shap_interaction}. These values help to explain not only the individual impact of a feature on the prediction but also how pairs of features jointly contribute beyond the sum of their individual effects, namely
\begin{equation}
    \hspace{-1cm}\phi_{i,j}=\sum_{S \subseteq N \setminus \{i\}} \frac{|S|!(N-|S|-2)!}{2(N-1)!} \left[f(S \cup \{i,j\}) + f(S) - f(S \cup \{i\}) - f(S \cup \{j\}) \right].\label{eq:shap_int}
\end{equation}
Following similar lines of reasoning outlined in the previous section on ALE, the total prediction $f(x)$ can be decomposed into a sum of all feature contributions plus a constant offset $f_0$. This is given by:
\begin{equation}
    f(x) \approx f_0 + \sum_{i=1}^{n} \phi_i + \sum_{1\leq i< j \leq n}^n \phi_{i,j}. \label{eq:shap}
\end{equation}
Conventionally, SHAP estimates yield local explanations (i.e. based on a specific instance of inputs). However, these can be further aggregated to provide global explainability of the model. Note that this follows a similar form to ALE in eq.(\ref{eq:ale}) which enables us to parametrise the aggregated SHAP predictions. In an analogous way to ALE, SHAP analysis can be used to detect interaction effects between input features which can be fed back to construct a better input feature dataset and improve the predictive power of the machine learning model \cite{shap_nature, shap_interaction}. Lastly, similarly to ALE, SHAP is also a model agnostic approach which makes it suitable for applications to a large variety of ``black-box'' machine learning models \cite{explainable_ai_book_2, explainable_ai_book_1}.

\subsection{Rationale of combining ALE and SHAP}
In this work the ALE and SHAP estimates are derived from the training dataset which was originally used to train the models. There are several reasons behinds this rationale. The training dataset is significantly larger and more diverse than the test dataset or the data contained in an individual plasma pulse. This enables us to understand what feature relationships, their importance and interactions the models have learned and most importantly, to what extent those relationships are general across the different plasma scenarios. Both ALE and SHAP approaches share common properties in that they can help explain the behaviour of input features over a given range, their importance, as well as detect their interactions. Applying these two fundamentally different approaches to a ``black-box'' model to explain its behaviour serves as a critical cross-validation mechanism helping us to independently verify the model behaviour. Furthermore, since both approaches are model-agnostic, this presents an additional degree of freedom for corroboration. By training multiple machine learning models using different prediction architectures and training methodologies, we can assess whether these various models demonstrate convergence. Put differently, we can inquire whether the models have acquired the same understanding of the dataset and exhibit a consistent set of principles governing their predictions. This, however, may not be the case in most applications due to the fact that a significant portion of the problems are ill-defined and do not have a unique solution. As a result, the model convergence and generalisations may not agree. These last points are particularly important to consider when monitoring the model performance in deployment. Understanding model decisions on an individual and global scale on unseen data can help corroborate whether the models generalise well and help identify problems with input features \cite{shap_nature}. Another implication in using such models is in plasma control, where the controller must satisfy stability and convergence properties to be used safely \cite{plasma_control}. The stability and convergence properties cannot be reliably computed for ``black-box'' models. However, model parametrisation using the interpretability tools coupled with the Hoeffding-Sobol decompoosition would permit such analysis. 

\section{Results and discussion}

\subsection{Model interpretability}
\begin{figure*}[t!]
\vspace{-1.cm}
\includegraphics[width=\textwidth]{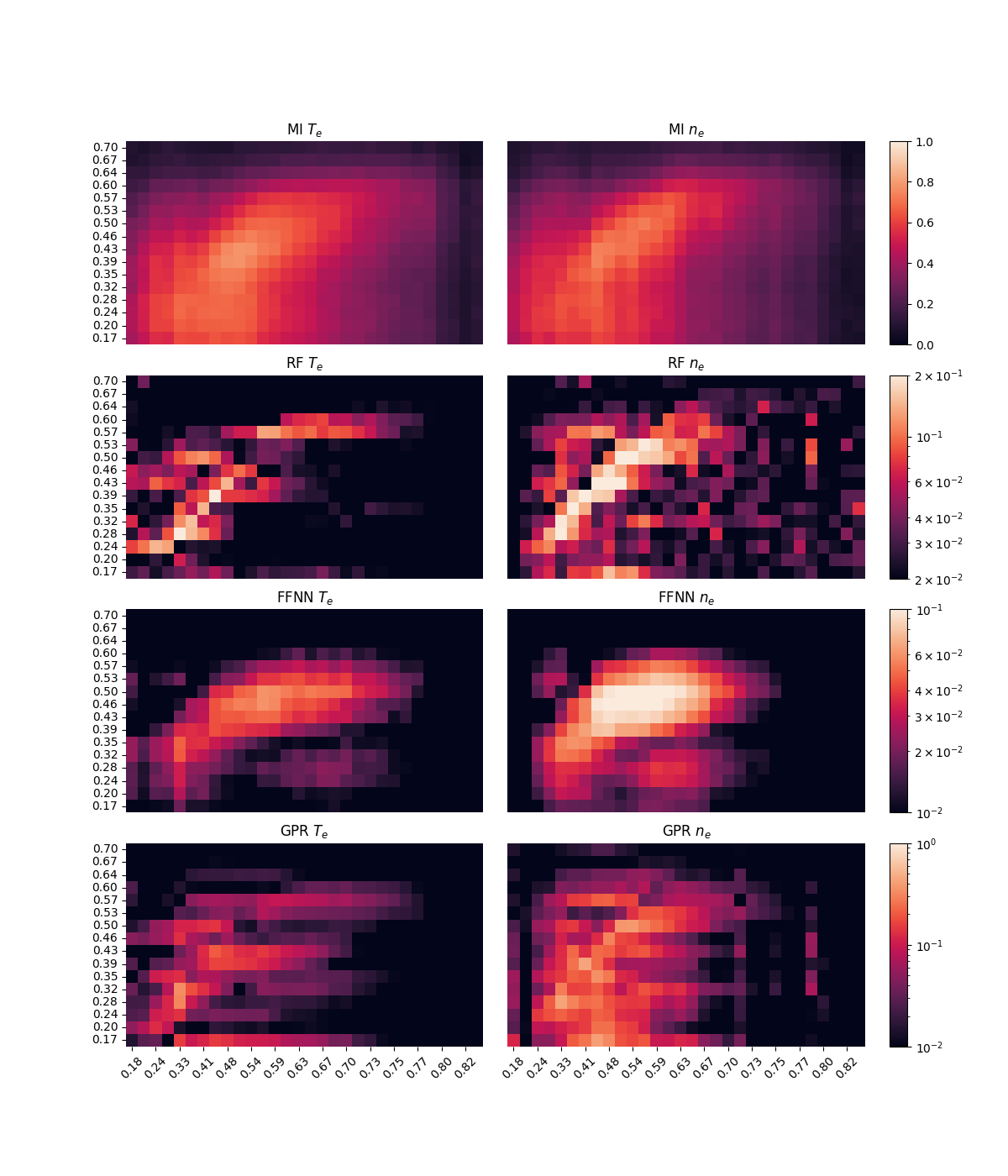}
\put(-420,170){\rotatebox{90}{SXR line of sight impact radius [m]}}
\put(-350,25){Thomson Scattering line of sight radial position [m]}
\put(-405,463){\small(a)}
\put(-225,463){\small(b)}
\put(-405,358){\small(c)}
\put(-225,358){\small(d)}
\put(-405,254){\small(e)}
\put(-225,254){\small(f)}
\put(-405,151){\small(g)}
\put(-225,151){\small(h)}
\vspace{-1cm}
\caption{(a)-(b) Mutual Information heatmap scores between measured soft X-ray emission for each line of sight signals and Thomson Scattering temperature and density lines of sight calculated using eq.~\ref{eq:mi}. High MI scores imply high degree of dependence between input - output variables. (c)-(h) RF, FFNN and GPR $T_e$ and $n_e$ are calculated mean absolute ALE contributions of measured SXR line of sight emissivities to predicted temperature/density of a given TS line of sight for a Random Forest, Feed-Forward Neural Network and Gaussian Process Regression models respectively. Here the heatmap units for MI are dimensionless. For $T_e$ and $n_e$ the heatmap units are in [keV] and [$m^{-3}/10^{20}$] respectively.}\label{fig:mae_ale}
\end{figure*}

We begin our model interpretability process by computing mean absolute values of Accumulated Local Effects (ALE) and visualise them as a heatmap plot for each trained model (see Figure~\ref{fig:mae_ale}). This provides us with basic understanding how each input feature in the model contributes (in absolute terms) towards the output predictions and therefore gives us first hand information on input feature importance and their distribution. We additionally show heatmaps of Mutual Information (MI) estimates calculated using eq.~\ref{eq:mi} between our inputs and output. The MI estimates being model independent are used as a first stage consistency check of feature importance and cross-validation against the computed ALEs. High mutual information scores indicate high mutual dependence between the input-output variables. In a magnetically-confined plasma, despite initial expectations of zero mutual information (MI) between soft X-ray (SXR) measurements at large impact parameters and Thomson scattering (TS) points at low radii (R), there is a correlation due to the flux surface structure and the limitations on the gradients the plasma can support. This results in the MI map being symmetric about the diagonal, rather than resembling a lower triangle. \\ \\
Consistency between high MI and ALE model estimates indicates that the models are using the input features for predictions effectively. Simultaneously, we also observe regions where mean absolute ALE estimates appear to be high (e.g. Figure~\ref{fig:mae_ale} Random Forest (RF), Gaussian Process Regression (GPR) $n_e$ SXRC line of sight 9, 10 and 12 at TS line of sight 4 which corresponds to the left hand side of the heatmap where SXR $R= \{0.32-0.6\}m$ and TS $R=0.78m$) - a behaviour that is much less pronounced in MI scores or Feed-Forward Neural Network (FFNN) ALE estimates. This could indicate potential model overfitting or an outlier in ALE estimation which drives a higher mean absolute ALE score. For most of the plasma scenarios included in this work, the radial positions of TS $R>0.78$~m do not collect much scattered light as the plasma is not large enough. As a result, the measured TS temperatures and densities in those regions are very close to zero making them very weakly correlated with the SXR emission measurements. As such, we do not see significant contribution from the SXR measurements towards the predicted temperature and density in these regions of the heatmap.
\begin{figure*}[t!]
\vspace{-1.cm}
\includegraphics[width=\textwidth]{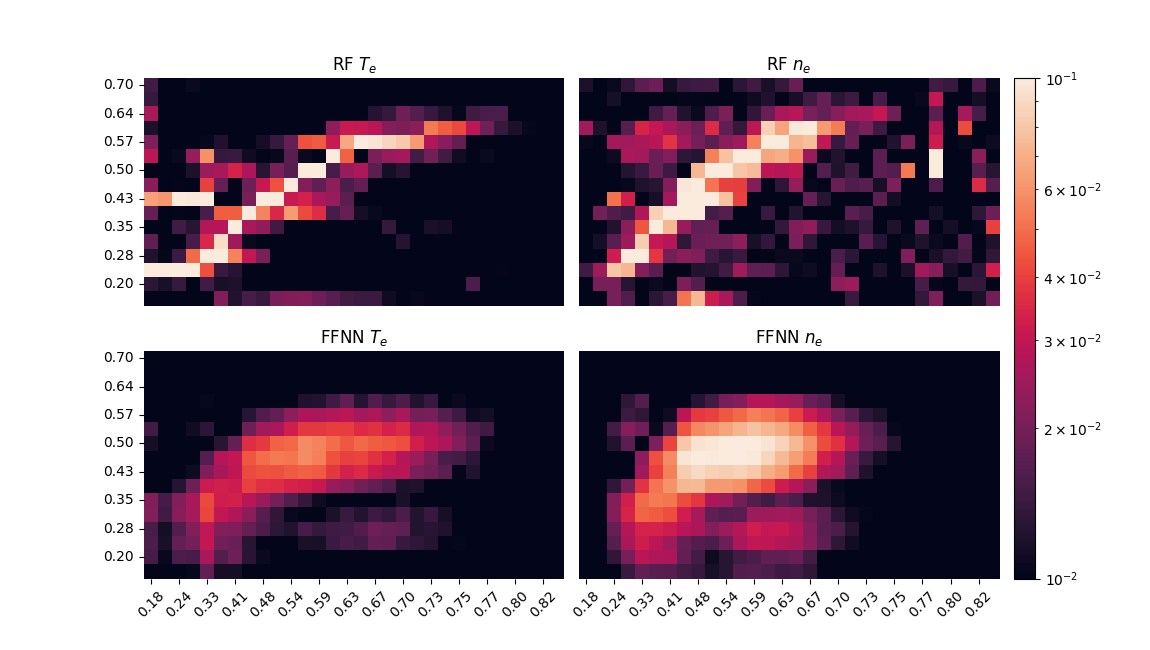}
\put(-405,228){\small(a)}
\put(-225,228){\small(b)}
\put(-405,120){\small(c)}
\put(-225,120){\small(d)}
\put(-420,40){\rotatebox{90}{SXR line of sight impact radius [m]}}
\put(-350,0){Thomson Scattering line of sight radial position [m]}
\caption{Mean absolute SHAP contributions of measured SXRC line of sight emissivities to predicted temperature/density of a given TS line of sight for a Random Forest (RF) (a)-(b) and Feed-Forward Neural Network (FFNN) (c)-(d) models. For $T_e$ and $n_e$ the heatmap units are in [keV] and [$m^{-3}/10^{20}$] respectively. Due to the fact that SHAP calculations are generally computationally intense, especially for large models, we were not able to perform them on our GPR model.}\label{fig:mae_shap}
\end{figure*}
\hspace{-0.15cm}From the estimated mean-absolute ALE scores we observe clear visual differences in the heatmap signatures within each model (i.e. whether the prediction is temperature or density) as well as between the models. The RF (Random Forest) model shows a noisy diagonal distribution between the SXRC emission line of sight and its contribution towards the predicted temperature or density. While the overlap between the SXRC and TS lines of sight in the heatmap data aligns somewhat with physical expectations, spurious contributions in $n_e$ appear around TS lines of sight $R\gtrapprox 0.8m$ where scattered signals are not typically expected. \\ \\
In contrast, the FFNN (Feed Forward Neural Network) heatmap is more consistent with the fact that the measured SXRC emission is line-integrated, ``seeing'' different regions of the plasma and covering different output lines of sight. The RF model appears to provide a more localised contribution from the SXRC lines of sight whilst suppressing the contributions from other SXRC lines of sight that have overlapping regions when viewed in the y-x plane. The GPR heatmaps display an approximate intermediate case between the RF and FFNN model heatmaps where the mean-absolute input contributions towards the predictions are diffuse (similar to FFNN) with a degree of specificity similar to RF (e.g. SXRC line of sight impact radius $R=0.57$~m contributions across different TS lines of sight for $T_e$). \\ \\
In an analogous fashion to ALE, we compute mean absolute contributions with SHAP (see Figure~\ref{fig:mae_shap}). Unlike ALE, which estimates the average local contribution of a feature to the predicted output across a set of observations, SHAP can quantify the specific contribution of each feature for individual predictions. As a result, SHAP estimates can reveal feature contributions that would otherwise be averaged-out in ALE estimates. However, since we are focusing on the mean field contributions we would not expect to see a significant difference in the computed heatmaps between ALE and SHAP. Indeed comparing Figures~\ref{fig:mae_ale} and \ref{fig:mae_shap} we observe near-identical behaviour between mean absolute ALE and SHAP estimates for FFNN model. Our FFNN architecture is reasonably simple (2 deep layers). This normally translates to having smoothly distributed feature effects. In contrast to this, the RF model is likely capturing much more complex feature behvaiour through ensemble tree structures which result in highly localised feature effects as seen in SHAP estimates. Moreover, having ReLU activation function in our FFNN results in smooth transitions (unlike having discrete and sharp transitions in RFs) which likely results in more consistent and stable feature attributions across different methods. Due to the fact that SHAP calculations are generally computationally intense, especially for large models, we were not able to perform them on our GPR model. \\ \\
\begin{figure*}[t!]
\includegraphics[width=\textwidth]{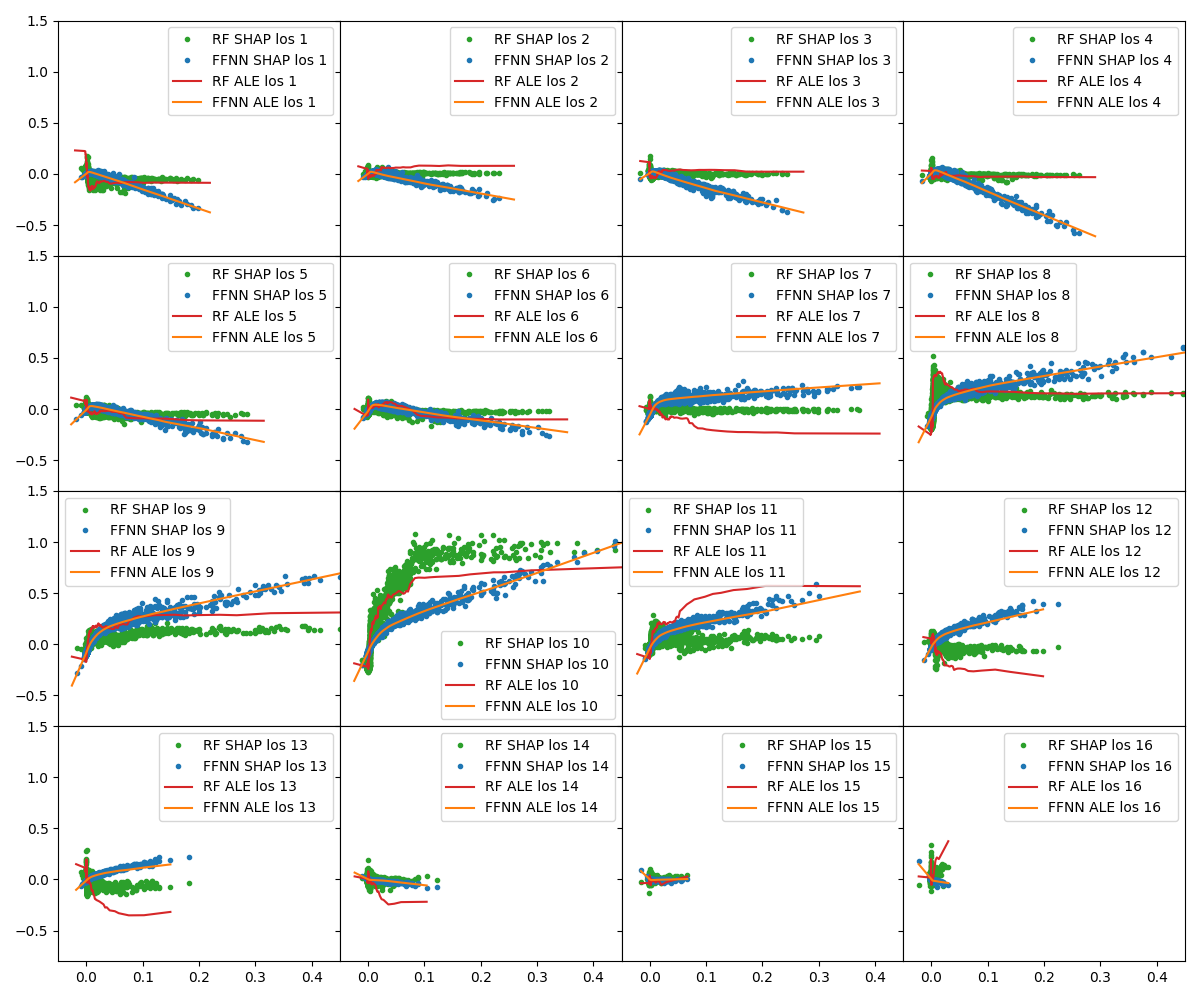}
\put(-460,70){\rotatebox{90}{$\delta T_e(R=0.52$~m) [keV] contribution from input feature }}
\put(-300,-10){Normalised soft X-ray emission $\epsilon$}
\caption{\label{fig:shap_ale_te} Computed Accumulated Local Effects and SHAP values for each input emission of the soft X-ray signal using the training dataset for radial output line of sight $R=0.52$~m of an FFNN and RF multi-output models. The y-axis corresponds to fractional contribution to the predicted output temperature as a function of the normalised soft X-ray emission amplitude. The lines of sight 1-6 correspond to high-field side whereas lines of sight 13-16 correspond to the low field side. See figure~\ref{fig:diagnostic_layout} for additional layout details. The corresponding density interpretability plot from the same model is shown in the appendix in figure~\ref{fig:shap_ale_ne_appendix}.}
\end{figure*}
\hspace*{-0.24cm}The heatmap estimates show a global picture of intepretability between input and output features for a given model i.e. which SXRC emission input channel contributes to which output/s of the predicted electron temperature and density averaged over the channel emission. However, this mean-field approach fails to convey how the amplitude of the measured emission influences the predicted output of a given line of sight. To visualise this, we take a vertical cross section of the heatmap (without averaging for a given output temperature/density line of sight and plot the measured emission amplitude on the x-axis against the computed ALE or SHAP values for a multi-output RF and FFNN models, see Figure~\ref{fig:shap_ale_te} (additional figures \ref{fig:shap_ale_ne_appendix}-\ref{fig:ale_te_model_comparison_appendix} in the appendix). The dependence plots in the figure correspond to $ALE(\epsilon_i)$ and $\phi(\epsilon_i)$ i.e. accumulated local effect (eq.~\ref{eq:ale_i}) and SHAP (eq.~\ref{eq:shap_i}) contributions to the predicted temperature/density as a function of emission $\epsilon_i$ for a given SXRC line of sight $i$. The ALE estimates are averaged over a specified emission interval $l$ or region whilst the SHAP estimates are computed for each individual set of observations. The mean-absolute heatmap plots of ALE and SHAP yielded near-identical signatures for FFNN multi-output model. This behaviour is further reflected in the dependence plots in figures~\ref{fig:shap_ale_te} (and \ref{fig:shap_ale_ne_appendix} in the appendix) where we observe good correspondence between the different interpretability methods for the multi-output FFNN model. The multi-output RF model interpretability shows lower fidelity between the two different techniques which was already seen in the heatmap plots. However, this picture provides a more detailed breakdown of this calculated divergence. Here, the RF model ALE-SHAP behaviour has an approximate feature correspondence between the curves which is offset by some constant. The RF ALE curves show additional irregular structure in their estimates which are most likely mediated by the discrete decision architecture of the random forest model. Furthermore, this fine structure can be averaged out or enhanced by adjusting the interval length $l$ in eq.~\ref{eq:ale_i}. Whilst SHAP was too expensive to compute for our multi-output GPR model, we were able to obtain GPR ALE estimates (see figures~\ref{fig:ale_ne_model_comparison_appendix} and \ref{fig:ale_te_model_comparison_appendix} of GPR benchmarked against FFNN and RF in the appendix). In contrast to RF and FFNN ALE behaviour the GPR ALE estimates convey richer and more complex behaviour. Yet, as observed in figures~\ref{fig:model_predictions}-\ref{fig:interferometer_comparison} all models show comparable predictive power as well as similar MAE performance (see table~\ref{tab:mae_performance}). \\ \\
To test the robustness of the model interpretability results we performed hyperparameter tuning on the machine learning models (focusing on single output model predictions instead of 30 pairs for the same 16 inputs). Bayesian and Hyperband optimisers were used for RF and FFNN models respectively \cite{bo, hyperband}. Model optimisation was found to have modest improvements in MAE scores (compared to the untuned models as baseline). More importantly, we found the ALE / SHAP estimates on fine tuned models did not change significantly to affect their interpretability. The line shapes and their structure from the dependency plots remained near identical, with small vertical offsets and marginal line gradient changes. \\ \\
So far in our discussion we have neglected the pair-wise input feature interaction terms described in equations~\ref{eq:ale} and \ref{eq:shap_int}). Figures~\ref{fig:shap_ale_te} and \ref{fig:shap_ale_ne_appendix} in the appendix) showing SHAP estimates against emission contain non-negligible vertical dispersion. Typically vertical dispersion is associated with interaction between input features \cite{shap_nature}. This will be discussed in the next section. \\ \\
The calculated ALE and SHAP dependence plots convey a mapping between the input and output features $f\colon \epsilon \rightarrow T_e, n_e$ where the measured emission for a given line of sight is given in eq.~\ref{eq:emission}. This emission function has complex dependence on detector characteristics, impurity content and integrated field of view. However, the results obtained from the dependence plots for FFNN multi-output model show a rather simple and human interpretable picture of how the ``black-box'' model is transforming the measured data to make new predictions. It has been shown that deep neural networks have strong inductive bias towards simple functions for input-output mapping (low Kolmogorov complexity) \cite{nn_generalisation_1, nn_generalisation_2}. Our interpretability estimates corroborate this result for our neural network model. These properties further align with the modern interpretation of the scientific method where model simplicity is associated with higher empirical content and model falsifiability \cite{popper}. Here simplicity is interpreted as equivalent to the paucity of adjustable parameters in the model. Having said that, these statements alone are not enough to justify the model selection criteria and need to be further scrutinised via empirical tests in order to falsify them. This will be discussed in the last section.
  
\subsection{Model compression/parametrisation through interpretability}
The calculated mapping between input features and output predictions of our FFNN model is dominated by its individual input feature contributions with the respective relationships given by the ALE and SHAP dependency plots in figures~\ref{fig:shap_ale_te} and \ref{fig:shap_ale_ne_appendix}. Using Hoeffding-Sobol decomposition outlined in the previous section we can parameterise the input-output relationship to generate a ``grey-box'' model. Here the choice of the term ``grey-box'' implies that we know the approximate functional form of model mapping, but we do not necessarily have a full understanding of the physics driving those predictions. In contrast, a ``white-box'' model would provide a complete picture of the underlying physics derived from its underlying assumptions. \\ \\
Figure~\ref{fig:parametrised_predictions} (and \ref{fig:parametrised_predictions_appendix} in the appendix) show parametrised FFNN model derived from ALE and SHAP estimates compared against the original ``black-box'' model predictions and the measured TS temperature and density based on Hoeffding-Sobol (HS) decomposition. In this particular case we neglect first order pair-wise interactions. For SHAP HS decomposition we perform symbolic regression on the SHAP dependency plots to obtain a generalised parametric equation that estimates electron density and temperature for a given line of sight of the following form \cite{sr,py_sr}
\begin{figure*}[t!]
\includegraphics[width=0.95\textwidth]{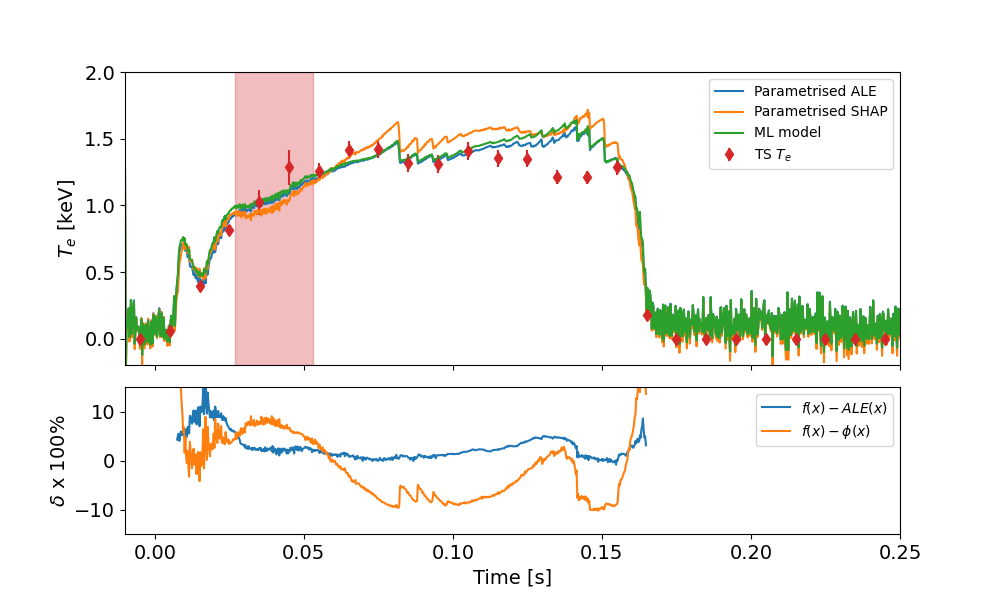}
\put(-410,220){(a)}
\put(-410,85){(b)}
\caption{\label{fig:parametrised_predictions} ST40 plasma pulse 11574. (a) Time dependent temperature trace for radial line of sight $R=0.53$~m. The shaded red region indicates where the TS beam/light collection optics suffer from transient pointing instability due to mechanical vibrations. The diamond tickers show the measured temperature by TS. The solid lines show the predicted time dependent density of the same TS line of sight using the Hoeffding-Sobol decomposed ALE and SHAP models as well as the original FFNN model. (b) Calculated percentage relative error time evolution between the original ML model and the Hoeffding-Sobol decomposed ALE and SHAP models.}
\end{figure*}
\begin{equation}
    n_e(\epsilon)=\sum_{i=1}^{n}a_i \epsilon_i + \frac{g_i\epsilon_i}{\epsilon_i+c_i}+ b_i,\label{eq:pysr_ne}
\end{equation}
\begin{equation}
    T_e(\epsilon)=\sum_{i=1}^{n} \alpha_i \epsilon_i + \frac{\gamma_i\epsilon_i}{\epsilon_i+\theta_i} + \beta_i,\label{eq:pysr_te}
\end{equation}
where $\epsilon_i$ is the emission amplitude for a given line of sight $i$ and $a_i, b_i$ etc., are constants derived from symbolic regression fitting. For the ALE dependency plots we perform simple interpolation to create a lookup table which is then used to generate the predictions. Parametrisation or look-up table based model inference is easy to implement on hardware for real-time applications as it does not rely on model specific architectures for inference circumventing the need for bespoke model compilers \cite{keras2c, fastml_code}. Moreover, if the trained model has interpretability with negligible feature interaction effects, then the functional decomposition of the model is reduced to a linear combination of the functional forms of the ALE or SHAP dependency plots which have fewer parameters than the original model used for training. This means that we can compress the original model without a significant loss in performance in the predictions. Requiring fewer mathematical operations to perform predictions further reduces the loop-bandwidth time constraints relevant for real-time applications. \\ \\
As a benchmarking comparison, the ALE derived look-up model table for temperature (neglecting first order interaction effects) has a 3$\%$ higher mean-absolute relative error for the above example compared to the original model while the parametrised SHAP model based on fitted equations (\ref{eq:pysr_ne}$-$\ref{eq:pysr_te}) with 64 (16 SXR emission measurements with a maximum of 4 free parameters to fit) parameters yielding a mean absolute relative error of $6\%$ when compared to the original model. The original FFNN model consists of $\approx$3k trainable parameters. Similar values are observed for the density predictions. The degree of model compression will inevitably depend on the complexity of the problem for which the model is being trained on. In our case, the first order interactions were comparatively smaller, however, if this was not the case then we would need to consider additional $C(16,2)=120$ pairwise interactions terms. We find that the contribution of pairwise interactions towards the decomposed FFNN model predictions are negligible. Similar trends are observed with the HS decomposed RF model via ALE look-up table and SHAP parametrisation. One key difference between FFNN and RF ALE estimates is that the latter show non-smooth behaviour which is then translated to non-smooth predictions in the HS decomposition. This is not the case for RF parametrised HS decomposition using symbolic regression. The remaining discrepancies between the original model and ALE/SHAP decomposed models can be attributed to the nature of ALE averaging effects as well as imperfect parametrisation function/fitting respectively. The ALE/SHAP interactions terms for our GPR model were prohibitively expensive to compute and therefore are not shown.  
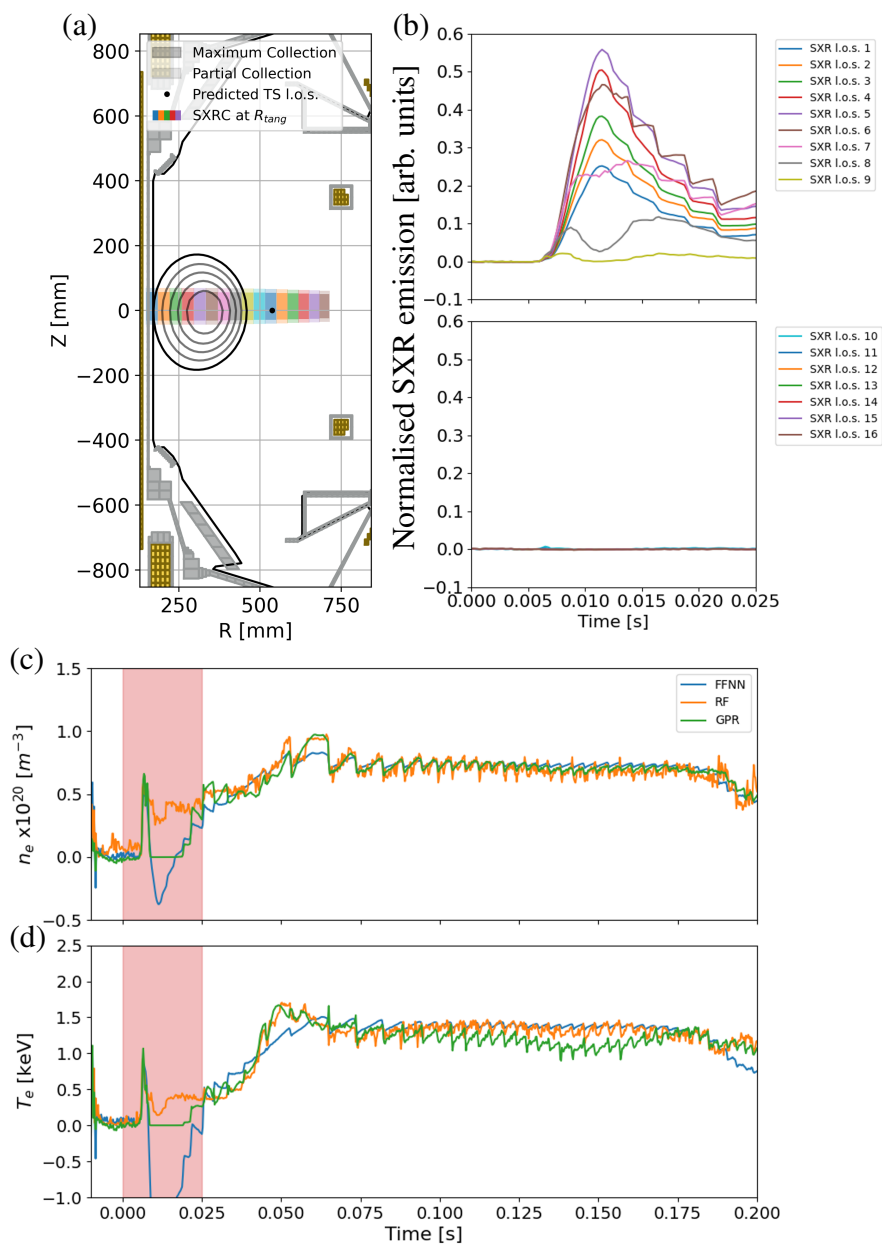
\begin{figure*}[t!]
\centering
\includegraphics[width=0.8\textwidth]{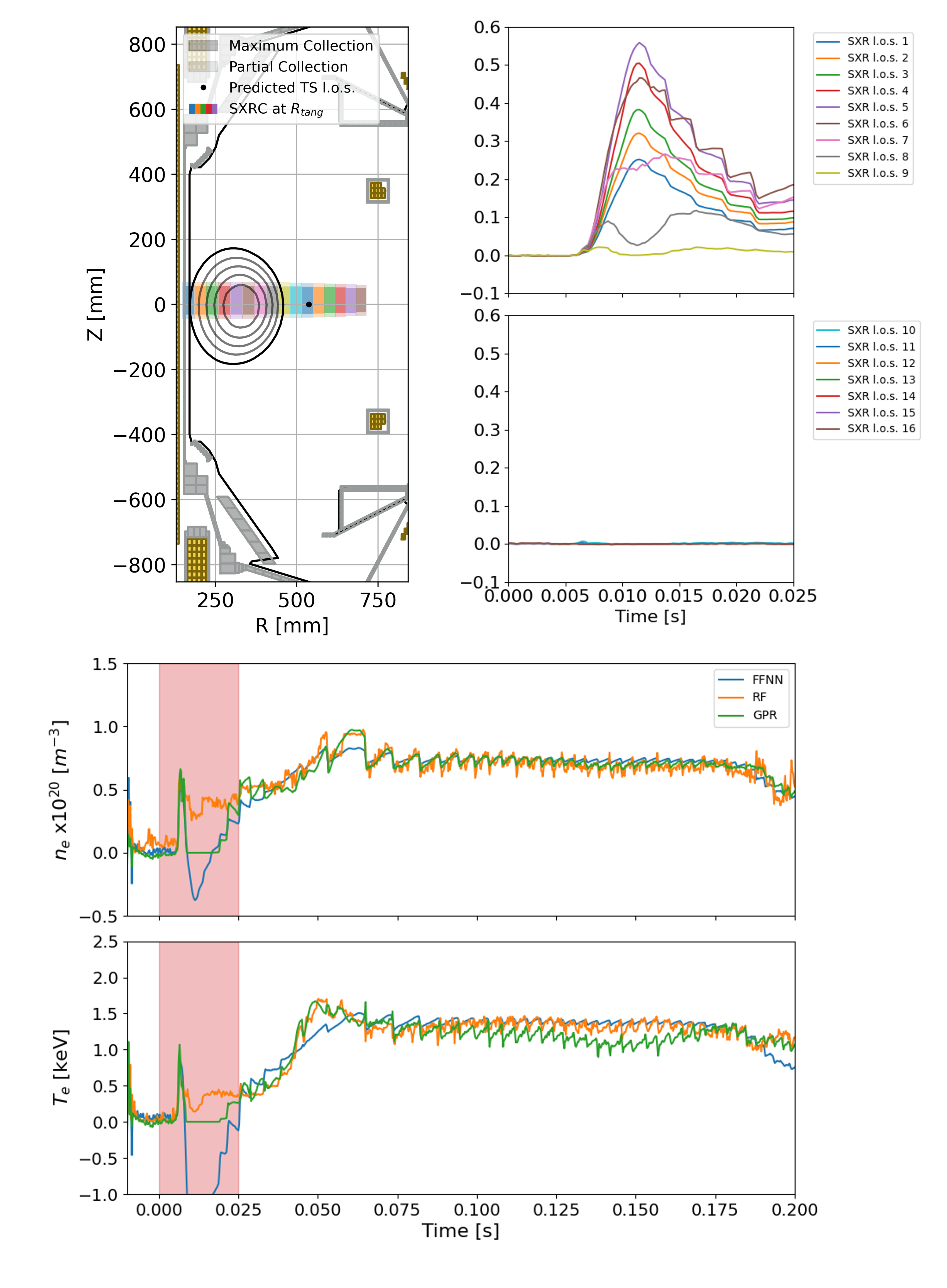}
\put(-320,475){(a)}
\put(-195,475){(b)}
\put(-340,235){(c)}
\put(-340,130){(d)}
\put(-195,280){\rotatebox{90}{Normalised SXR emission [arb. units]}}
\vspace{-0.7cm}
\caption{\label{fig:model_falsification} ST40 plasma pulse 12012. (a) magnetic reconstruction of the plasma at $t=0.012$~s (b) SXR signal traces corresponding to plasma region (lines of sight 1-9) and remaining lines of sight  10-16 where there is no plasma. (c) and (d) time evolution of the predicted density and temperature values for radial line of sight $R=0.53$~m corresponding to the black dot in (a). Here the shaded regions indicate where the FFNN predicts negative temperature/density. The colour coding of SXR lines of sight in (a) correspond to signal traces in (b).}
\end{figure*}


\subsection{Model falsifiability/reliability}
Now, we shift our focus to identifying instances where the models produce unphysical results. This analysis is required if we want to develop an understanding of the limitations of the predictions and, more importantly, the underlying reasons for these discrepancies. Additionally, this approach provides a direct comparison between the models, enabling us to assess whether a model aligns with physics-driven expectations.\\ \\ 
Consider the dependency plots shown in figure~\ref{fig:shap_ale_te} (and \ref{fig:shap_ale_ne_appendix} in the appendix). These plots reveal that the lines of sight on the high field side (SXRC lines of sight 1 - 6, as detailed in figure~\ref{fig:diagnostic_layout}) negatively influence the predicted temperature and density at the radial position $R=0.53$~m, which approximates the plasma core location. This effect is observed when analysing the emission data using the multi-output Feed-Forward Neural Network (FFNN) model. In contrast, the Random Forest (RF) model shows negligible sensitivity of the predicted temperature/density to changes in measured emission for these lines of sight, as indicated by a flat ALE/SHAP profile. The Gaussian Process Regression (GPR) model presents a more complex interaction (detailed in figures~\ref{fig:ale_ne_model_comparison_appendix} and \ref{fig:ale_te_model_comparison_appendix} in the appendix). In scenarios where the plasma is concentrated on the high-field side (a condition occasionally met during the startup phase), the measured SXR emission is expected to be significantly higher from lines of sight with $R<0.4$~m and markedly lower from the core region ($0.4<R<0.6$~m), where less plasma is present. Upon examining the dependency plots for $R=0.53$~m in figures~\ref{fig:shap_ale_te}-\ref{fig:shap_ale_ne_appendix} and \ref{fig:ale_ne_model_comparison_appendix}-\ref{fig:ale_te_model_comparison_appendix}, we would anticipate that the FFNN model would predict a negative temperature/density at this radial position originating from a dominant negative contribution from high-field side data points, as depicted in the ALE/SHAP plots. This contrasts with the RF and GPR models, where the ALE/SHAP estimates (ALE only for GPR) suggest the temperature at $R=0.53$~m should be approximately zero. Validating these models against the pulse 12012 data from the most recent campaign, which was not included in the model training dataset, reveals that the FFNN model indeed predicts a negative temperature at startup - a direct consequence of having a small plasma. This prediction is not mirrored by the RF or GPR models, as shown in figure~\ref{fig:model_falsification}. RF predicts a small positive temperature/density whilst GPR  predicts zero. This further allows us to falsify the RF model and corroborate the GPR predictions with physics driven expectations. Consequently, the analysis presented here demonstrates a process under which interpretability tools can be used to deduce the conditions for model falsification and help explain them. Contrasting this to a case of using a ``black-box'' model only, one would have to run many different scenarios to ascertain the underlying reasons for its predictions.

\section{Conclusion}
In this work, we presented a machine learning workflow that constructs a ``grey-box'' synthetic diagnostic, leveraging the key advantages of individual diagnostics while avoiding their intrinsic limitations. By utilising state-of-the-art interpretable machine learning tools, we demonstrated that it is possible to deconstruct a ``black-box'' model into an intuitive format, providing both local and global explanations of the model’s behavior and testing these explanations against physical expectations. Moreover, by employing various machine learning architectures and interpretability tools, we showed that despite similar performance in predictions, different models can yield different interpretations. Crucially, in the example demonstrated in this work, this enables us to construct a physics driven adversarial scenario that allows us to falsify the models and understand their failure mechanism. \\ \\
In the application explored in this work, we found that under certain conditions where model interpretability is straightforward, it is possible to parametrise the model into a closed functional form or a look-up table. In our case this resulted in a significant model compression without substantial loss in prediction accuracy. This lossy compression is advantageous for deploying machine learning models in real-time applications on hardware, as it eliminates the need for ML model-specific compilers and enables model-agnostic real-time applications. \\ \\
Maintaining plasma stability and mitigating disruptions are critical challenges in fusion research, where machine learning has shown promising results \cite{fpga_ml, turbulence1, turbulence2, turbulence3}. Building on this success, the application of interpretability tools used in this work could be leveraged to understand the underlying drivers for plasma disruption and further identify strategies for its prevention as well us draw comparisons with alternative approaches \cite{turbulence4}. 
\section*{Acknowledgements}
We thank Peter Buxton, Dmitry Osin, Thomas O'Gorman, Benjamin Vincent and Vadim Nemytov for reading the manuscript and providing useful suggestions.





\section*{References}

\bibliographystyle{unsrt}  

\begin{thebibliography}{41}


\bibitem{plasma_data_science_review}
Rushil Anirudh, Rick Archibald, M. Salman Asif, Markus M. Becker, Sadruddin Benkadda, Peer-Timo Bremer, Rick H. S. Budé et. al., \textit{2022 Review of Data-Driven Plasma Science}, IEEE Transactions on Plasma Science ( Volume: 51, Issue: 7, July 2023)  \url{https://doi.org/10.1109/TPS.2023.3268170}

\bibitem{plasma_ml_and_bayesian_review}
A. Pavone,1, A. Merlo, S. Kwak and J. Svensson, \textit{Machine learning and Bayesian inference in nuclear fusion research: an overview}, 2023 Plasma Phys. Control. Fusion 65 053001, \url{https://doi.org/10.1088/1361-6587/acc60f}

\bibitem{pinns}
Raissi, Maziar, Paris Perdikaris, and George E. Karniadakis. \textit{Physics-informed neural networks: A deep learning framework for solving forward and inverse problems involving nonlinear partial differential equations}, Journal of Computational Physics 378 (2019): 686-707. \url{https://doi.org/10.1016/j.jcp.2018.10.045}

\bibitem{attempt_at_explainable_ai_for_plasma}
M. S. Parsons, \textit{Interpretation of machine-learning-based disruption models for plasma control}, 2017 Plasma Phys. Control. Fusion 59 085001, \url{https://doi.org/10.1088/1361-6587/aa72a3}

\bibitem{keras2c_paper}
Rory Conlina, Keith Ericksonb, Joeseph Abbatec, Egemen Kolemena, \textit{Keras2c: A library for converting Keras neural networks to real-time compatible C}, Engineering Applications of Artificial Intelligence, Volume 100, April 2021, 104182, \url{https://doi.org/10.1016/j.engappai.2021.104182}

\bibitem{keras2c}
\url{https://github.com/f0uriest/keras2c}

\bibitem{machine_learning_pcs}
Degrave, J., Felici, F., Buchli, J. et al., \textit{Magnetic control of tokamak plasmas through deep reinforcement learning}, Nature 602, 414–419 (2022). \url{https://doi.org/10.1038/s41586-021-04301-9}

\bibitem{rt_profile_prediction}
F. Felici et. al., \textit{Real-time-capable prediction of temperature and density profiles in a
tokamak using RAPTOR and a first-principle-based transport model}, 2018 Nucl. Fusion 58 096006, \url{https://doi.org/10.1088/1741-4326/aac8f0}

\bibitem{fpga_ml}
Y. Wei, R. F. Forelli, C. Hansen, et.al., \textit{Low latency optical-based mode tracking with machine learning deployed on FPGAs on a tokamak}, Rev. Sci. Instrum. 95, 073509 (2024) \url{https://doi.org/10.1063/5.0190354}

\bibitem{synthetic_diagnostic_example_1}
C. M. Samuell, A. G. Mclean, C. A. Johnson, F. Glass, and A. E. Jaervinen, \textit{Measuring the electron temperature and identifying plasma detachment using machine learning and spectroscopy }, Rev. Sci. Instrum. 92, 043520 (2021), \url{https://doi.org/10.1063/5.0034552}

\bibitem{synthetic_diagnostic_example_2}
Bishop, C., Strachan, I., O’Rourke, J., Maddison, G., \& Thomas, P. (1993). \textit{Reconstruction of tokamak density profiles using feedforward networks}, Neural Computing \& Applications, 1(1), 4–16. \url{https://doi.org/10.1007/BF01411370}

\bibitem{synthetic_diagnostic_example_3}
D. J. Clayton, K. Tritz, D. Stutman, R. E. Bell, A. Diallo, B. P. LeBlanc and M. Podest\`a, \textit{Electron temperature profile reconstructions from multi-energy SXR measurements using neural networks}, Plasma Phys. Control. Fusion 55 (2013) 095015 (8pp), \url{https://doi.org/10.1088/0741-3335/55/9/095015}

\bibitem{synthetic_diagnostic_example_4}
Jonathan Chalaturnyk and Richard Marchand, \textit{A First Assessment of a Regression-Based Interpretation of Langmuir Probe Measurements},  Front. Phys., 03 May 2019, Sec. Space Physics, \url{ https://doi.org/10.3389/fphy.2019.00063}

\bibitem{sxr_temperature_multi_energy}
Tkachenko, E.E., Kurskiev, G.S., Zhiltsov, N.S. et al. \emph{Application of Machine Learning to Determine Electron Temperature in Globus-M2 Tokamak Using the Soft X-Ray Emission Data and the Thomson Scattering Diagnostics Data}. Phys. Atom. Nuclei 85, 1214–1222 (2022). \url{https://doi.org/10.1134/S1063778822070122}

\bibitem{sxr_temperature_reconstruction_analytic}
P. Franz; F. Bonomo; L. Marrelli; P. Martin; P. Piovesan; G. Spizzo; B. E. Chapman; D. Craig; D. J. Den Hartog; J. A. Goetz; R. O’Connell; S. C. Prager; M. Reyfman; J. S. Sarff, \textit{Two-dimensional time resolved measurements of the electron temperature in MST}, Rev. Sci. Instrum. 77, 10F318 (2006), \url{https://doi.org/10.1063/1.2229192}

\bibitem{ale_main}
Daniel W. Apley, Jingyu Zhu, \textit{Visualizing the Effects of Predictor Variables in Black Box Supervised Learning Models}, (2016), arXiv:1612.08468, \url{https://doi.org/10.48550/arXiv.1612.08468}

\bibitem{ale_py}
Daniel W. Apley, Jingyu Zhu, \textit{Accumulated Local Effects (ALE) and Package ALEPlot} (2018), \url{https://www.google.com/url?sa=t&source=web&rct=j&opi=89978449&url=https://cran.r-project.org/web/packages/ALEPlot/ALEPlot.pdf&ved=2ahUKEwju35nMnryFAxUSVEEAHS3yBWwQFnoECBEQAQ&usg=AOvVaw10VM4SLPfTrXSrB9VNVbLA}

\bibitem{ale_code}
\url{https://github.com/DanaJomar/PyALE}

\bibitem{alibi}
Janis Klaise and Arnaud Van Looveren and Giovanni Vacanti and Alexandru Coca, \textit{Alibi Explain: Algorithms for Explaining Machine Learning Models}, Journal of Machine Learning Research, (2021), 22, 181, \url{http://jmlr.org/papers/v22/21-0017.html}

\bibitem{alibi_git}
\url{https://github.com/SeldonIO/alibi/}

\bibitem{shap_original}
Shapley, L., 1953. \textit{A value for n-person games. In: Contrib. Theory of Games, II.} In: Ann. Math. Stud., vol. 28,
pp. 307–317. \url{https://doi.org/10.1515/9781400881970-018 }

\bibitem{shap_comp}
Scott M. Lundberg, Su-In Lee, \textit{A Unified Approach to Interpreting Model Predictions}, Advances in Neural Information Processing Systems 30 (NIPS 2017). \url{https://doi.org/10.48550/arXiv.1705.07874}

\bibitem{shap_code}
\url{https://github.com/shap/shap}

\bibitem{sr}
Miles Cranmer, \textit{Interpretable Machine Learning for Science with PySR and SymbolicRegression.jl}, (2023), arXiv:2305.01582, \url{https://doi.org/10.48550/arXiv.2305.01582}

\bibitem{py_sr}
\url{https://github.com/MilesCranmer/PySR}


\bibitem{st40}
S. A. M. McNamara, et.al., \textit{Achievement of ion temperatures in excess of 100 million degrees Kelvin in the compact high-field spherical tokamak ST40}, Nucl. Fusion 63 (2023) 054002 (6pp). \url{https://doi.org/10.1088/1741-4326/acbec8}

\bibitem{sxr_cary}
C. Colgan, et.al, \textit{Soft X-Ray Tomography on the High Field Spherical Tokamak ST40}, (2024).

\bibitem{putterich2019determination}
P{\"u}tterich et al., \textit{Determination of the tolerable impurity concentrations in a fusion reactor using a consistent set of cooling factors}, Nuclear Fusion 59 (5), (2019). \url{https://doi.org/10.1088/1741-4326/ab0384}


\bibitem{ts_hazel}
H. Lowe et al., \textit{First measurements of electron temperature and density profiles using Thomson scattering on the ST40 spherical tokamak} 20th International Symposium on Laser-Aided Plasma Diagnostics (2023).


\bibitem{tensorflow}
\url{https://github.com/tensorflow/tensorflow}

\bibitem{rf_gpr}
Olivier Grisel, Andreas Mueller, Lars, Alexandre Gramfort, Gilles Louppe, Thomas J. Fan, Peter Prettenhofer, et al. \textit{Scikit-learn/scikit-learn: Scikit-learn 1.5.0}. Zenodo, May 21, 2024. \url{https://doi.org/10.5281/zenodo.11237090}.

\bibitem{fastml_paper}
Duarte, Javier and others, \textit{Fast inference of deep neural networks in FPGAs for particle physics}, 2018 JINST 13 P07027, \url{https://doi.org/10.1088/1748-0221/13/07/P07027}

\bibitem{fastml_code}
FastML Team, \textit{fastmachinelearning/hls4ml}, (2023), Zenodo, v0.8.1, github: \url{https://github.com/fastmachinelearning/hls4ml}, \url{https://doi.org/10.5281/zenodo.1201549}

\bibitem{real_time_control}
S. Joung et al., \textit{Deep neural network Grad–Shafranov solver constrained with measured magnetic signals}, Nucl. Fusion, vol. 60, no. 1, Jan. 2020, Art. no. 016034. \url{https://doi.org/10.1088/1741-4326/ab555f}

\bibitem{adam}
Diederik P. Kingma, Jimmy Ba, \textit{Adam: A Method for Stochastic Optimization}, (2014) \url{https://doi.org/10.48550/arXiv.1412.6980}

\bibitem{mutual_info}
G. Nicoletti and D. M. Busiello, \textit{Mutual Information Disentangles Interactions from Changing Environments}, Phys. Rev. Lett. 127, 228301, (2021), \url{https://doi.org/10.1103/PhysRevLett.127.228301}

\bibitem{mutual_information}
Alexander Kraskov, Harald St\"ogbauer, and Peter Grassberger, \textit{Estimating mutual information}, Phys. Rev. E 69, 066138 – Published 23 June 2004, \url{https://doi.org/10.1103/PhysRevE.69.066138}.

\bibitem{freedman_diaconis}
Freedman, David, Diaconis, Persi. \textit{On the histogram as a density estimator: L2 theory}, Probability Theory and Related Fields. 57 (4): 453–476. (1981) \url{https://doi.org/10.1007/BF01025868} 

\bibitem{explainable_ai_book_2}
Christoph Molnar, \textit{Interpretable Machine Learning
A Guide for Making Black Box Models Explainable}, (2023). \url{https://christophm.github.io/interpretable-ml-book/}

\bibitem{anova_0}
G. E. B. Archer , A. Saltelli and I. M. Sobol  \textit{Sensitivity
measures,anova-like Techniques and the use of bootstrap}, Journal of Statistical
Computation and Simulation, 58:2, 99-120, (1997) \url{https://doi.org/10.1080/00949659708811825}

\bibitem{anova_1}
B. Efron, C. Stein , \textit{The Jackknife Estimate of Variance}, Ann. Statist. 9(3): 586-596 (May, 1981). \url{https://doi.org/10.1214/aos/1176345462}. 

\bibitem{hoeffding_sobol}
G. Chastaing, F. Gamboa, C. Prieur, \textit{Generalized Hoeffding-Sobol Decomposition for Dependent Variables -Application to Sensitivity Analysis}, (2011), arXiv:1112.1788, \url{https://doi.org/10.48550/arXiv.1112.1788}

\bibitem{rabitz}
H. Rabitz, O.F. Alis, K. Shim, J. \textit{Shorter, Managing the tyranny of parameters in mathematical modelling}, in: Proceedings of SAMO’98, Venice, 1998, p. 209.

\bibitem{shap_nature}
Lundberg, S.M., Erion, G., Chen, H. et al. \textit{From local explanations to global understanding with explainable AI for trees}, Nat. Mach. Intell. \textbf{2}, 56–67 (2020). \url{https://doi.org/10.1038/s42256-019-0138-9}

\bibitem{shap_interaction}
Fujimoto, K., Kojadinovic, I. \& Marichal, J.-L. \textit{Axiomatic characterizations of probabilistic and cardinal-probabilistic interaction indices}. Games Econ. Behav. 55, 72–99 (2006). \url{https://doi.org/10.1016/j.geb.2005.03.002}

\bibitem{explainable_ai_book_1}
Holzinger, A., Goebel, R., Fong, R., Moon, T., Müller, KR., Samek, W. (2022). xxAI - Beyond Explainable Artificial Intelligence. In: Holzinger, A., Goebel, R., Fong, R., Moon, T., Müller, KR., Samek, W. (eds) \textit{xxAI - Beyond Explainable AI}, (2020), Lecture Notes in Computer Science, vol 13200. Springer, Cham. \url{https://doi.org/10.1007/978-3-031-04083-2_1}

\bibitem{plasma_control}
Karl Johan Åström and Richard Murray, \textit{Feedback Systems: An Introduction for Scientists and Engineers}, ISBN: 9780691193984, (2021).

\bibitem{bo}
Tim Head, MechCoder, Gilles Louppe, Iaroslav Shcherbatyi, fcharras, Zé Vinícius, cmmalone, Christopher Schröder, nel215, Nuno Campos, Todd Young, Stefano Cereda, Thomas Fan, rene-rex, Kejia (KJ) Shi, Justus Schwabedal, carlosdanielcsantos, Hvass-Labs, Mikhail Pak, Alexander Fabisch. (2018). scikit-optimize/scikit-optimize: v0.5.2 (v0.5.2). Zenodo. \url{https://doi.org/10.5281/zenodo.1207017}

\bibitem{hyperband}
O'Malley, Tom and Bursztein, Elie and Long, James and Chollet, Fran\c{c}ois and Jin, Haifeng and Invernizzi, Luca, \textit{Keras Tuner}, (2019), \url{https://github.com/keras-team/keras-tuner}

\bibitem{nn_generalisation_1}
Guillermo Valle-Pérez, Chico Q. Camargo, Ard A. Louis, \textit{Deep learning generalizes because the parameter-function map is biased towards simple functions}, Published as a conference paper at ICLR 2019, \url{https://doi.org/10.48550/arXiv.1805.08522}

\bibitem{nn_generalisation_2}
Chris Mingard, Henry Rees, Guillermo Valle-Pérez, Ard A. Louis, \textit{Do deep neural networks have an inbuilt Occam's razor?}, 
\url{https://doi.org/10.48550/arXiv.2304.06670}

\bibitem{popper}
K. Popper, \emph{The Logic of Scientific Discovery}, Routledge, 2nd edition, ISBN: 9780415278447, (21 Feb. 2002). 

\bibitem{turbulence1}
Semin Joung, David R Smith, G McKee, Z Yan, K Gill, J Zimmerman, B Geiger, R Coffee, FH O’Shea, A Jalalvand, E Kolemen, \emph{Tokamak edge localized mode onset prediction with deep neural network and pedestal turbulence}, Nucl. Fusion 64 (2024) 066038 (14pp), \url{https://doi.org/10.1088/1741-4326/ad43fb}

\bibitem{turbulence2}
Jaemin Seo, SangKyeun Kim, Azarakhsh Jalalvand, Rory Conlin, Andrew Rothstein, Joseph Abbate, Keith Erickson, Josiah Wai, Ricardo Shousha, Egemen Kolemen, \emph{Avoiding fusion plasma tearing instability with deep reinforcement learning}, Nature volume 626, pages 746–751 (2024), \url{https://doi.org/10.1038/s41586-024-07024-9 }

\bibitem{turbulence3}
Vega, J., Murari, A., Dormido-Canto, S. et al. \textit{Disruption prediction with artificial intelligence techniques in tokamak plasmas}. Nat. Phys. 18, 741–750 (2022). \url{https://doi.org/10.1038/s41567-022-01602-2}


\bibitem{turbulence4}
S. A. Sabbagh, J. W. Berkery, Y. S. Park, et. al., \textit{ Disruption event characterization and forecasting in tokamaks}, Phys. Plasmas 30, 032506 (2023), \url{ https://doi.org/10.1063/5.0133825}















\end{thebibliography}

\appendix
\newpage
\section{Additional Figures}
\begin{figure*}[h]
\vspace{-0.5cm}
\centering
\includegraphics[width=0.8\textwidth]{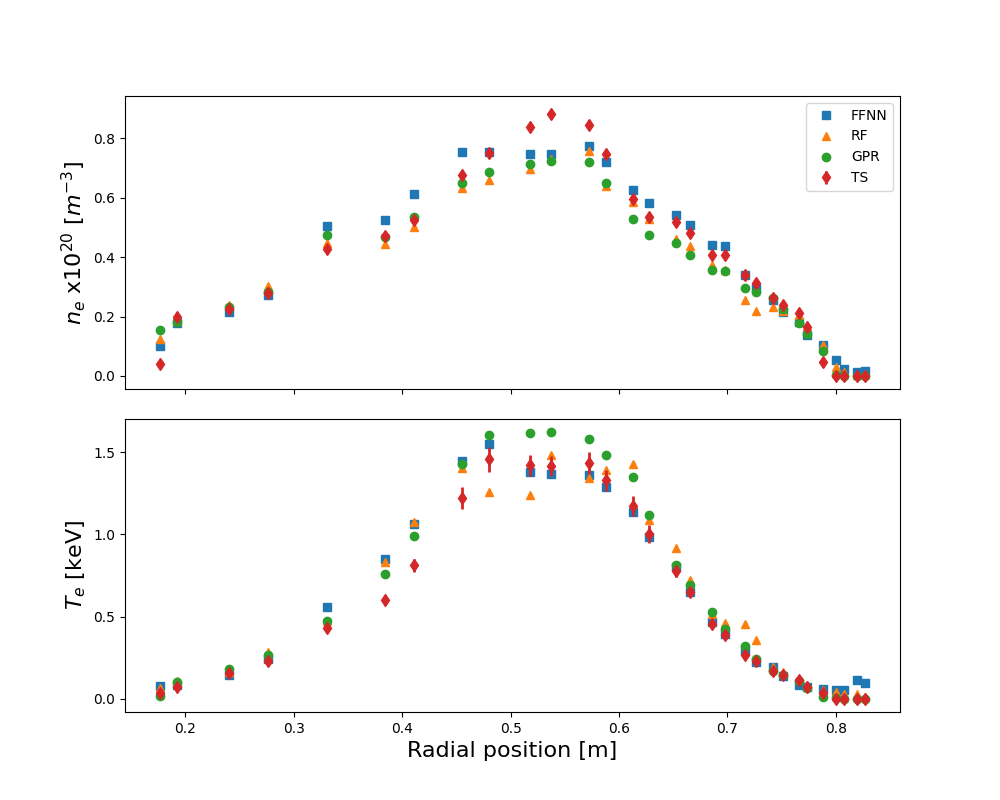}
\put(-350,250){(a)}
\put(-350,135){(b)}
\vspace{-0.5cm}
\caption{\label{fig:model_profile_predictions} ST40 plasma pulse 11574. Thomson Scattering profiles at 75ms mark for (a) density and (b) temperature as predicted by the different ML models benchmarked against measurements made by the Thomson Scattering diagnostic.}
\end{figure*}

\begin{figure*}[t!]
\includegraphics[width=\textwidth]{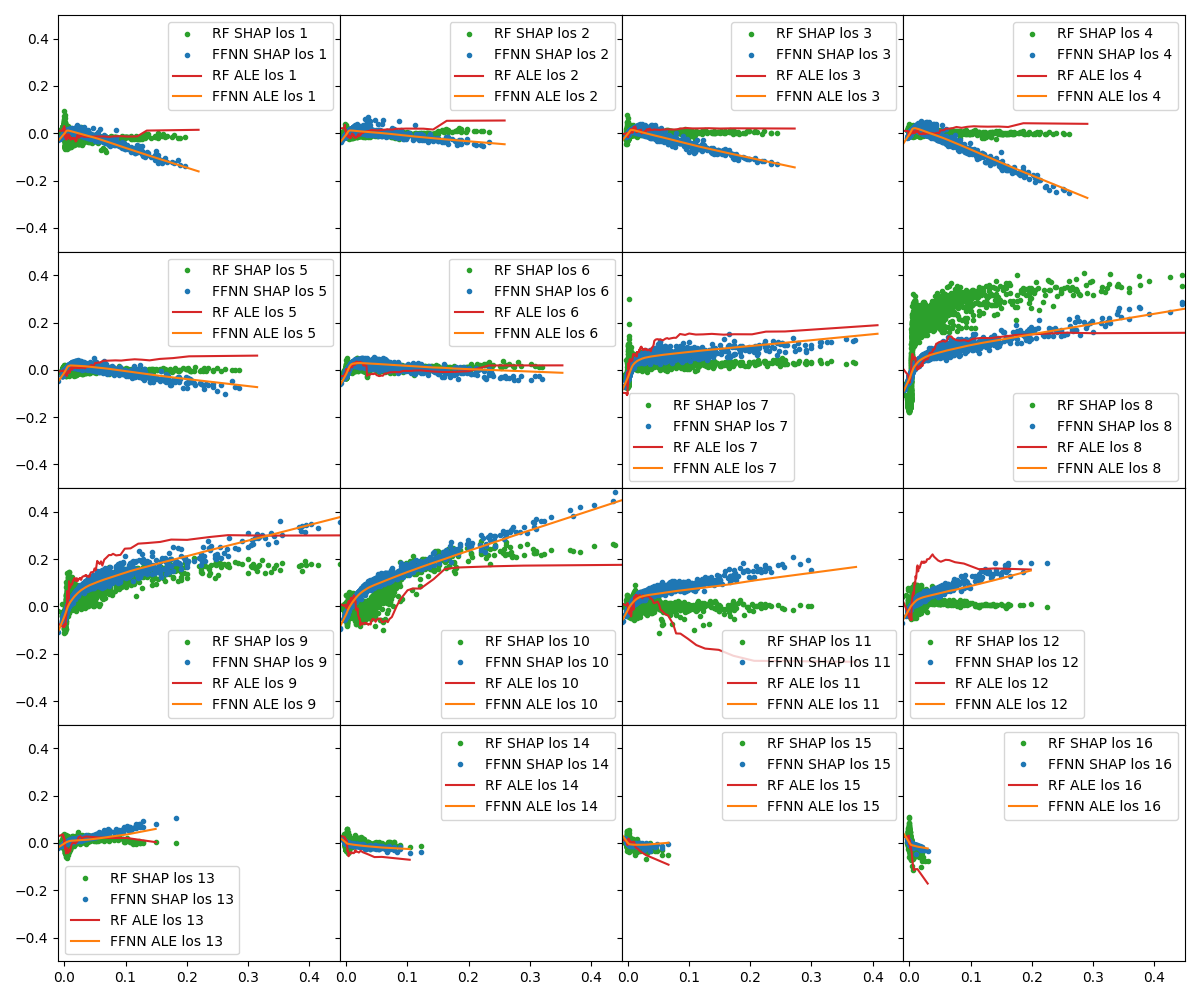}
\put(-460,60){\rotatebox{90}{$\delta n_e(\textrm{l.o.s. 20})\times 10^{20}$ [$m^{-3}$]  contribution from input feature }}
\put(-300,-10){Normalised soft X-ray emission $\epsilon$}
\caption{\label{fig:shap_ale_ne_appendix} Computed Accumulated Local Effects and SHAP values for each input emission of the soft X-ray signal using the training dataset for radial output line of sight $R=0.52$~m of an FFNN and RF multi-output models. The y-axis corresponds to fractional contribution to the predicted output density as a function of the normalised soft X-ray emission amplitude. The lines of sight 1-6 correspond to high-field side whereas lines of sight 13-16 correspond to the low field side. See figure~\ref{fig:diagnostic_layout} for additional layout details.}
\end{figure*}

\begin{figure*}[t!]
\includegraphics[width=\textwidth]{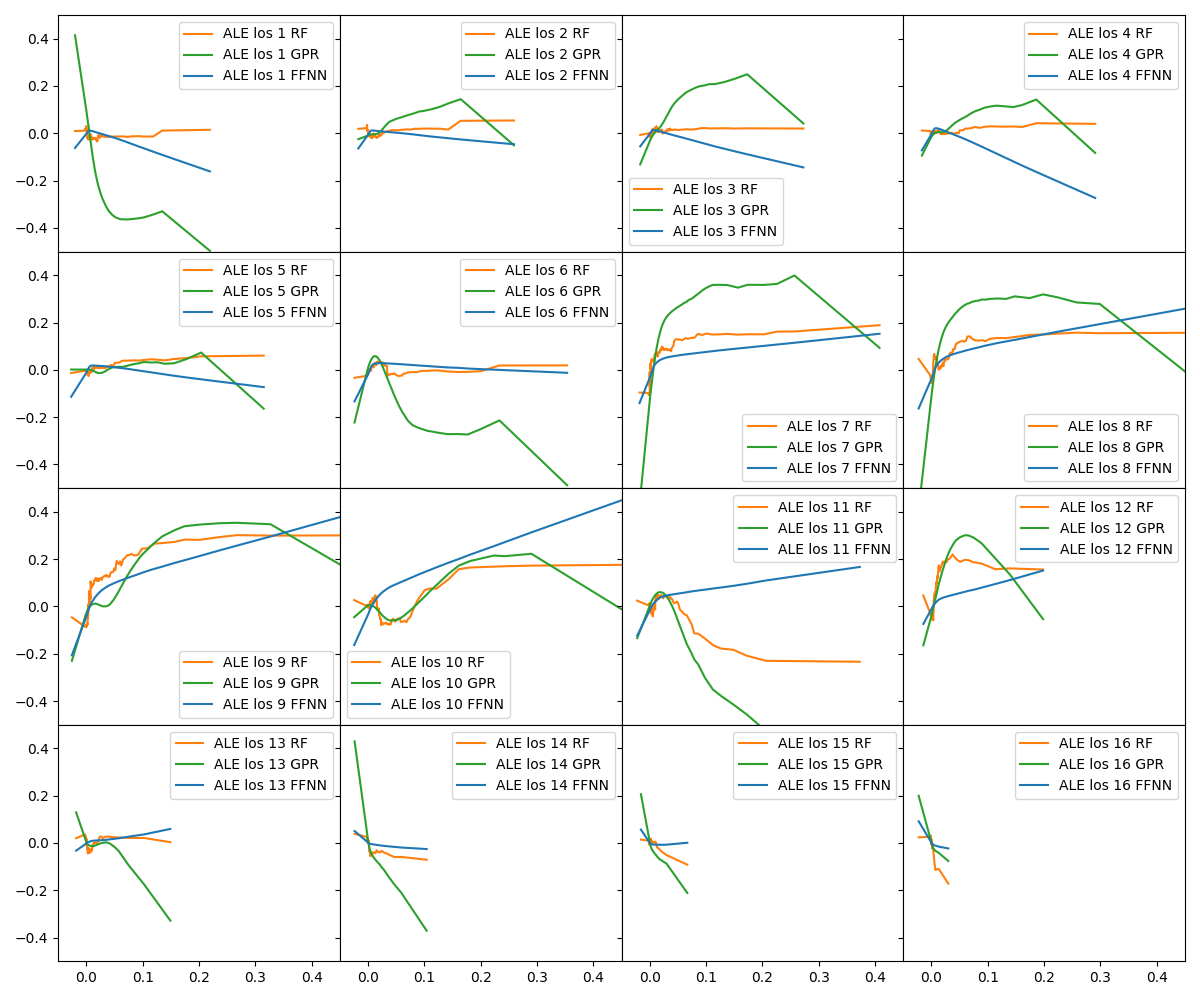}
\put(-460,60){\rotatebox{90}{$\delta n_e(\textrm{l.o.s. 20})\times 10^{20}$ [$m^{-3}$]  contribution from input feature }}
\put(-300,-10){Normalised soft X-ray emission $\epsilon$}
\caption{\label{fig:ale_ne_model_comparison_appendix} Computed Accumulated Local Effects for each input emission of the soft X-ray signal using the training dataset for radial line of sight $R=0.53$~m for multi-output models. The y-axis corresponds to fractional contribution to the predicted output density as a function of the input soft X-ray signal amplitude. The lines of sight 1-6 correspond to high-field side whereas lines of sight 13-16 correspond to the low field side. See figure~\ref{fig:diagnostic_layout} for additional layout details.}
\end{figure*}

\begin{figure*}[t!]
\hspace*{-1.8cm}
\includegraphics[width=1.2\textwidth]{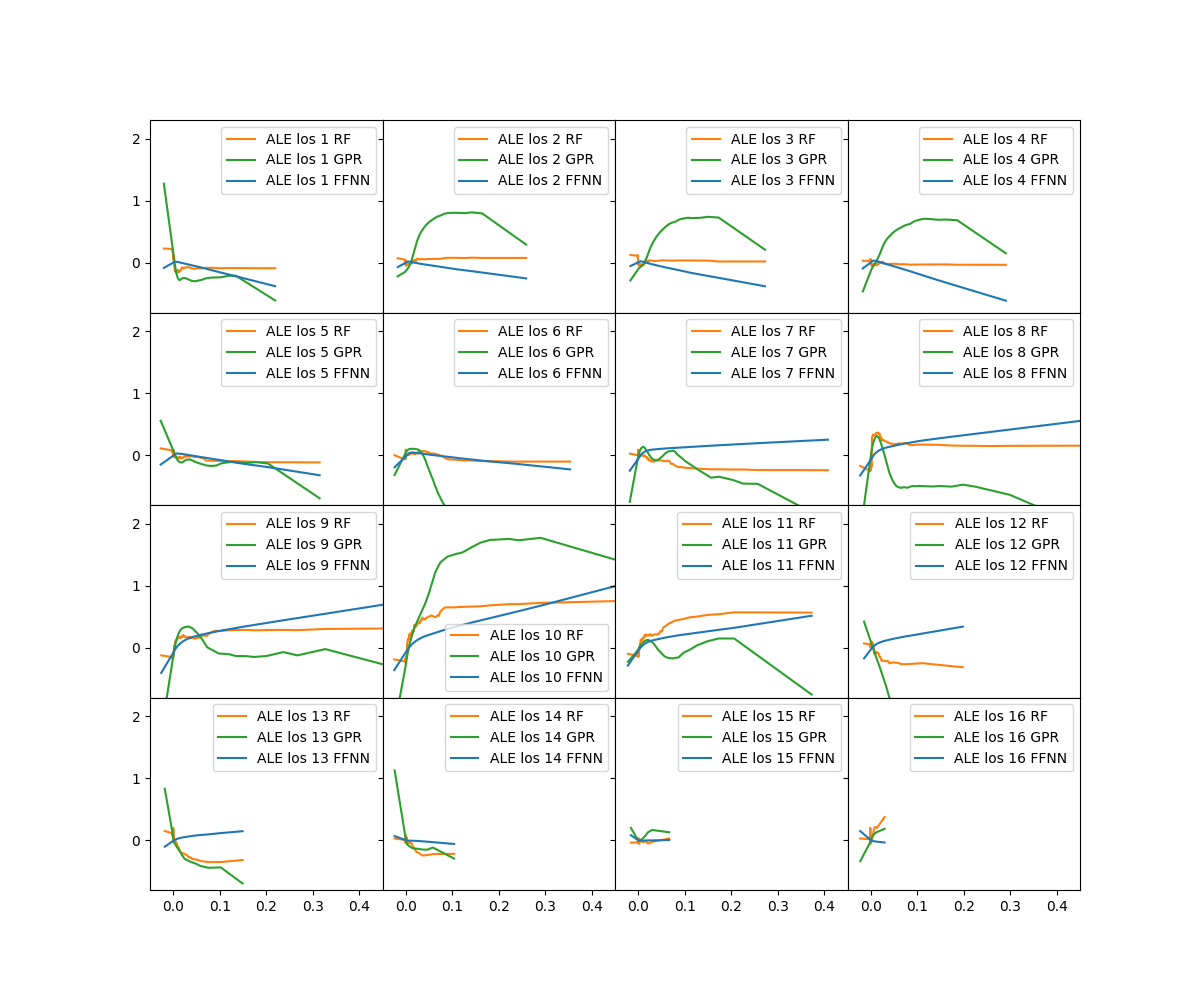}
\put(-500,90){\rotatebox{90}{$\delta T_e(R=0.52$~m) [keV] contribution from input feature }}
\put(-340,20){Normalised soft X-ray emission $\epsilon$}
\vspace{-1cm}
\caption{\label{fig:ale_te_model_comparison_appendix} Computed Accumulated Local Effects for each input emission of the soft X-ray signal using the training dataset for radial line of sight $R=0.53$~m for multi-output models. The y-axis corresponds to fractional contribution to the predicted output temperature as a function of the input soft X-ray signal amplitude. The lines of sight 1-6 correspond to high-field side whereas lines of sight 13-16 correspond to the low field side. See figure~\ref{fig:diagnostic_layout} for additional layout details.}
\end{figure*}

\begin{figure*}[t!]
\includegraphics[width=0.95\textwidth]{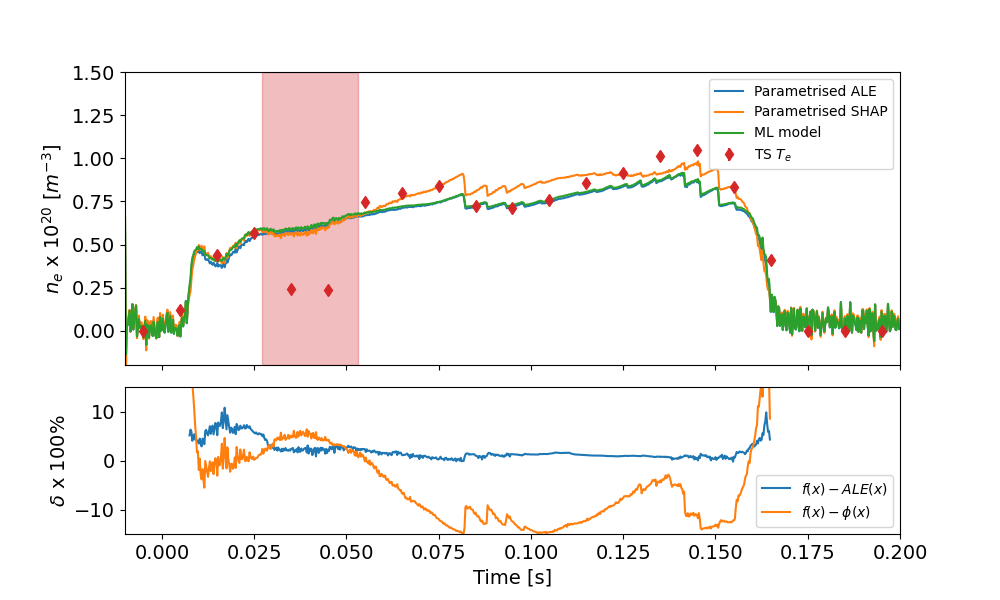}
\put(-410,220){(a)}
\put(-410,85){(b)}
\caption{\label{fig:parametrised_predictions_appendix} ST40 plasma pulse 11574. (a) Time dependent density trace for radial line of sight $R=0.53$~m. The shaded red region indicates where the TS beam/light collection optics suffer from transient pointing instability due to mechanical vibrations. The diamond tickers show the measured Thomson scattering density. The solid lines show the predicted time dependent density of the same TS line of sight using the Hoeffding-Sobol decomposed ALE and SHAP models as well as the original FFNN model. (b) Calculated percentage relative error time evolution between the original ML model and the Hoeffding-Sobol decomposed ALE and SHAP models.}
\end{figure*}


\end{document}